\documentclass[fleqn,usenatbib]{mnras}

\usepackage[T1]{fontenc}
\usepackage{ae,aecompl}

\usepackage{newtxtext,newtxmath}

\usepackage{graphicx}	% Including figure files
\usepackage{amsmath}	% Advanced maths commands
\usepackage{amssymb}	% Extra maths symbols\
\usepackage[shortcuts]{extdash}
\title[K2 observations of the Pleiades]{Beyond the \emph{Kepler}/K2 bright limit: variability in the seven brightest members of the Pleiades}

\author[T. R. White et al.]
{T. R. White,$^{1,2,3}$\thanks{E-mail: tim@phys.au.dk}
B.~J.~S.~Pope,$^{4}$
V.~Antoci,${^1}$
P.~I.~P\'apics,$^{5}$
C.~Aerts,$^{5,6}$
D.~R.~Gies,$^{7}$
\newauthor
K.~Gordon,$^{7}$
D.~Huber,$^{8,9,1,10}$
G.~H.~Schaefer,$^{11}$
S.~Aigrain,$^{4}$,
S.~Albrecht,$^{1}$
\newauthor
T.~Barclay,$^{12,13}$
G.~Barentsen,$^{12,13}$
P.~G.~Beck,$^{14}$
T.~R.~Bedding,$^{9,1}$
\newauthor
M.~Fredslund~Andersen,$^{1}$
F.~Grundahl,$^{1}$
S.~B.~Howell,$^{12}$
M.~J.~Ireland,$^{15}$
\newauthor
S.~J.~Murphy,$^{9,1}$
M.~B.~Nielsen,$^{16,3}$
V.~Silva~Aguirre,$^{1}$
P.~G.~Tuthill$^{9}$
\\
$^{1}$Stellar Astrophysics Centre, Department of Physics and Astronomy, Aarhus University, Ny Munkegade 120, DK-8000 Aarhus C, Denmark\\
$^{2}$Institut f\"{u}r Astrophysik, Georg-August-Universit\"{a}t G\"{o}ttingen, Friedrich-Hund-Platz 1, 37077 G\"{o}ttingen, Germany\\
$^{3}$Max-Planck-Institut f\"ur Sonnensystemforschung, Justus-von-Liebig-Weg 3, 37077 G\"ottingen, Germany\\
$^{4}$Oxford Astrophysics, University of Oxford, Denys Wilkinson Building, Keble Rd, Oxford OX1 3RH, UK\\
$^{5}$Instituut voor Sterrenkunde, KU Leuven, Celestijnenlaan 200D, 3001 Leuven, Belgium\\
$^{6}$Department of Astrophysics, IMAPP, Radboud University Nijmegen, PO Box 9010, 6500 GL Nijmegen, The Netherlands\\
$^{7}$Center for High Angular Resolution Astronomy and Department of Physics and Astronomy, Georgia State University, P.~O.~Box~5060, Atlanta, GA \\30302-410, USA\\
$^{8}$Institute for Astronomy, University of Hawai`i, 2680 Woodlawn Drive, Honolulu, HI 96822, USA\\
$^{9}$Sydney Institute for Astronomy (SIfA), School of Physics, University of Sydney, NSW 2006, Australia\\
$^{10}$SETI Institute, 189 Bernardo Avenue, Mountain View, CA 94043, USA\\
$^{11}$The CHARA Array of Georgia State University, Mount Wilson Observatory, Mount Wilson, CA 91023, USA\\
$^{12}$NASA Ames Research Center, Moffett Field, CA 94035, USA\\
$^{13}$Bay Area Environmental Research Inst., 560 Third St., West Sonoma, CA 95476, USA\\
$^{14}$Laboratoire AIM, CEA/DRF - CNRS - Univ. Paris Diderot - IRFU/SAp, Centre de Saclay, 91191 Gif-sur-Yvette Cedex, France\\
$^{15}$Research School of Astronomy \& Astrophysics, Australian National University, Canberra, ACT 2611, Australia\\
$^{16}$Center for Space Science, NYUAD Institute, New York University Abu Dhabi, P.O. Box 129188, Abu Dhabi, UAE
}

\date{Accepted XXX. Received YYY; in original form ZZZ}

\pubyear{2017}

\begin{document}
\label{firstpage}
\pagerange{\pageref{firstpage}--\pageref{lastpage}}
\maketitle

\begin{abstract}
The most powerful tests of stellar models come from the brightest stars in the sky, for which complementary techniques, such as astrometry, asteroseismology, spectroscopy, and interferometry can be combined. The K2 Mission is providing a unique opportunity to obtain high-precision photometric time series for bright stars along the ecliptic. However, bright targets require a large number of pixels to capture the entirety of the stellar flux, and bandwidth restrictions limit the number and brightness of stars that can be observed. To overcome this, we have developed a new photometric technique, that we call halo photometry, to observe very bright stars using a limited number of pixels. Halo photometry is simple, fast and does not require extensive pixel allocation, and will allow us to use K2 and other photometric missions, such as TESS, to observe very bright stars for asteroseismology and to search for transiting exoplanets. We apply this method to the seven brightest stars in the Pleiades open cluster. Each star exhibits variability; six of the stars show what are most-likely slowly pulsating B-star (SPB) pulsations, with amplitudes ranging from 20 to 2000\,ppm. For the star Maia, we demonstrate the utility of combining K2 photometry with spectroscopy and interferometry to show that it is not a `Maia variable', and to establish that its variability is caused by rotational modulation of a large chemical spot on a 10\,d time scale.
\end{abstract}


%%%%%%%%%%%%%%%%% BODY OF PAPER %%%%%%%%%%%%%%%%%%

\section{Introduction}
The \textit{Kepler} mission \citep{borucki10}, which had the primary purpose of detecting Earth-like planets via the transit method, has also been a boon for stellar astrophysics by providing precise photometric light curves for studying stellar variability across the Hertzsprung-Russell (H-R) diagram. 
In particular, \textit{Kepler} has advanced our understanding of B-type stars building on earlier work with the MOST \citep[e.g.][]{walker05,aerts06,saio06} and CoRoT \citep[e.g.][]{huat09,neiner09,degroote09,degroote10} space telescopes. Slowly pulsating B (SPB) stars, first identified by \citet{waelkens91}, are of particular interest. They pulsate in high-order gravity modes excited by the $\kappa$-mechanism operating on the opacity-bump associated with iron-group elements \citep{dziembowski93,gautschy93}, and are sensitive probes of the physical properties of the stellar core \citep{miglio08}. They can therefore shed light on poorly-understood processes that take place around the core, including overshooting, diffusive mixing, and internal differential rotation. These processes have a substantial impact on stellar lifetimes. While \textit{Kepler} has revealed variability in dozens of B stars \citep{debosscher11,balona11,mcNamara12}, the greatest impact has stemmed not from the quantity of observed stars, but from the high quality of the long, nearly-uninterrupted time series. In particular, careful analysis of the oscillation spectra of SPB stars has revealed the fingerprints of internal rotation and mixing processes in the cores of these stars \citep{papics14,papics15,papics17}, enabling tests of prescriptions of mixing \citep{moravveji15,moravveji16} and the derivation of internal rotational profiles \citep{triana15}.

The nominal \textit{Kepler} mission ended when the loss of two reaction wheels meant the telescope could no longer maintain stable pointing in its original field. However, by aligning the telescope along its orbital plane, the disturbing torque of solar pressure on the roll axis is minimized. This allows for relatively stable pointing controlled periodically by thrusters about the roll axis, and with the remaining two reaction wheels for the other axes. With this method of operation, the K2 mission is conducting a series of approximately 80-day-long photometric observing campaigns in fields along the ecliptic \citep{howell14}, providing new opportunities for searching for exoplanets and investigating stellar variability across the H-R diagram.

% Figure 1
\begin{figure*}
	\includegraphics[width=1.4\columnwidth]{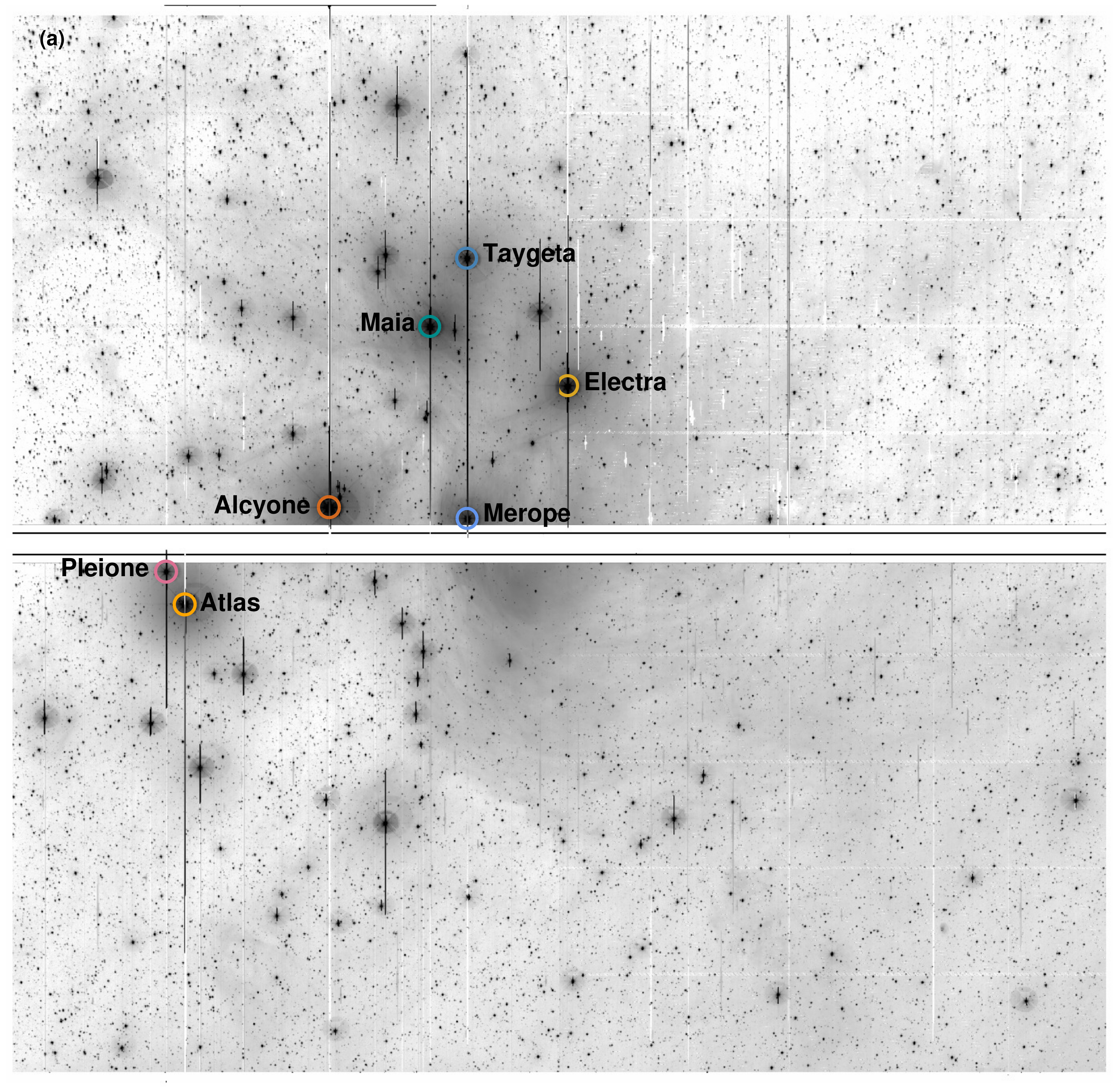}
	\includegraphics[width=0.68\columnwidth]{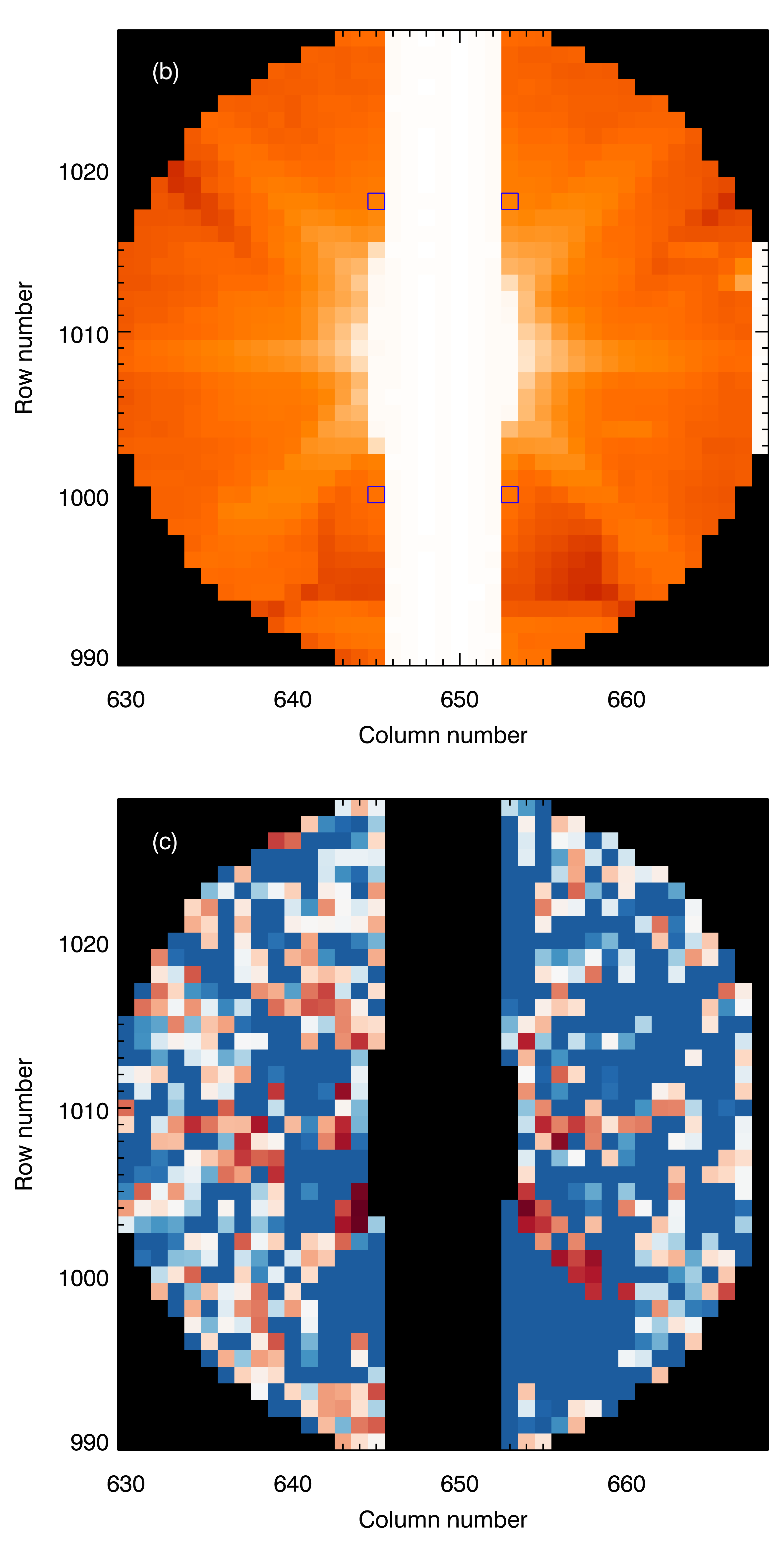}
    \caption{(a) Full frame image of Module~15 at the beginning of K2 Campaign 4. The target pixel masks around the seven brightest stars in the Pleiades are indicated: Alcyone (red), Atlas (orange), Electra (yellow), Maia (green), Merope (light blue), Taygeta (dark blue), and Pleione (pink). (b) Target pixel mask and \textit{Kepler} image of Alcyone. Segments of the light curves of the four highlighted pixels are shown in Fig.~\ref{fig:pixlc}. (c) Optimal weights for the Alcyone light curve. Red pixels have been up-weighted, while blue pixels have been down-weighted.}
    \label{fig:image}
\end{figure*}

The brightest stars ($V\lesssim6.0$\,mag) in the K2 fields are particularly desirable targets because these stars are the most amenable to complementary observations, including polarimetry and interferometry. 
The combination of asteroseismology with other modes of observation places strong constraints on stellar properties, by lifting degeneracies, and dependencies on stellar models \citep[e.g.][]{cunha07}. These well-characterized stars therefore provide the most stringent tests of stellar models, and can be used to examine internal processes \citep[e.g.][]{hjoerringgaard17}.

These bright stars also offer unique opportunities to precisely characterize transiting exoplanets and their atmospheres, as has been done with the well-studied 55~Cnc~e \citep[discovered to transit by][]{55cncdiscovery,55cncdiscovery2}, which orbits a $V=5.95$\,mag star. \citet{55cncmap} have shown the planet to have an extremely large temperature gradient and \citet{55cncvariable} have shown it displays variability in thermal emission from its dayside atmosphere. Only two known planets transit a brighter star, HD~219134 \citep[$V=5.57$;][]{motalebi,gillonmotalebi}, and adding even one new candidate to this list would be a significant breakthrough for exoplanetary science.

However, observations of bright stars present unique challenges. Most significantly, telemetric bandwidth limitations restrict the amount of data that can be downloaded from the \textit{Kepler} spacecraft. Pixels are only downloaded from `postage stamps' around pre-selected targets. The \textit{Kepler} CCDs saturate for stars with $Kp\lesssim$ 11--12\,mag, with excess flux bleeding along CCD columns. To recover the entirety of the stellar flux, the postage stamps for bright stars must include long bleed columns, thereby requiring a very large number of pixels. Consequently, data from few of the brightest stars are ordinarily selected to be downloaded. For the nominal \textit{Kepler} mission, the field was even chosen to avoid as many bright stars as possible to minimize the impact they would have on the primary, planet-finding mission \citep{koch10}. Of the 14\,stars on active silicon brighter than $Kp$\,=\,6\,mag, only $\theta$\,Cyg \citep{guzik16}, V380\,Cyg \citep{tkachenko12}, and 16\,Cyg\,A \citep{metcalfe12} were targeted over multiple observing quarters, HD\,185351 \citep{johnson14}, and HR\,7322 (Stokholm et al. in preparation) were only observed for a single quarter in short cadence mode (58.85\,s sampling), and nine stars were completely unobserved.

The desire to observe the brightest possible targets therefore calls for novel methods to recover light curves from an economically feasible number of pixels. \citet{pope16} recently showed that light curves of bright stars can be recovered from \textit{Kepler} calibration `smear' data, allowing for the observation of stars that were not specifically targeted. While a very valuable method and going a long way to solving this problem, it has a few drawbacks that lead us to continue to pursue other methods. One such drawback is the much shorter integration time and small number of pixels used for smear measurements, which leads to lower photometric precision with a photon noise equivalent to a star $\sim6.7\,\mathrm{mag}$ fainter. Another is that smear measurements are made along CCD columns, so targets falling on the same columns are confused. This is a particular concern in clusters, such as the Pleiades, and in the crowded K2 fields in the galactic plane. Finally, the bleed columns of bright stars, particularly those near the edge of the CCD, can also saturate the smear pixels in those columns, rendering them useless.

Another method has been developed by \citet{aerts17} to observe the O9.5Iab star HD\,188209 with \textit{Kepler}. This $V$\,=\,5.63\,mag supergiant was intentionally placed between CCDs during the nominal \textit{Kepler} mission. Nevertheless, scattered light from this bright star `contaminated' two nearby targets on active silicon, allowing for the variability in the supergiant to be successfully recovered.

We have developed this idea further with a new method that uses the flux recorded by non-saturated pixels in the halo surrounding a bright star. In this paper, we show that stellar variability may be successfully observed with this limited aperture, despite losing much of the stellar flux. While this method is susceptible to aperture losses, compounded by systematics inherent in the K2 mission, namely pointing drifts and inter- and intra-pixel sensitivity variation, we show that appropriate weighting of the contributions from each pixel can effectively neutralize these effects. To illustrate this method, we have applied it to the seven brightest stars in the Pleiades, all of which are late B-type stars, and were observed during K2~Campaign~4.

\section{Halo Photometry}

\subsection{Data}
K2 Campaign 4 observed a field in the direction of the constellation Taurus from 2015 February 8 to 2015 April 20. Fig.~\ref{fig:image}(a) shows the full-frame image of the Pleiades open cluster (M\,45) on the two CCDs of Module~15 of the \textit{Kepler} spacecraft. The seven brightest stars, ranging from $Kp$\,=\,2.99 to 5.19 mag \citep{huber16}, were each targeted with a circular aperture with a radius of 20 pixels, and observed in long cadence mode (29.4\,min sampling). Each aperture consists of 1245 pixels, an allocation equivalent to approximately six to twelve $Kp$\,=\,12\,mag stars. In contrast, a $Kp$\,=\,4\,mag star with a regular mask would require approximately $23400$ pixels to capture the entirety of its almost $1400$-pixel-long bleed column. Electra and Merope were near the edge of the CCD, and their target apertures were truncated along one side by several columns. Fainter cluster members were targeted with regular apertures. A detailed investigation of the rotation of these stars using K2\,data has been recently reported \citep{rebull16a,rebull16b,stauffer16}.
Fig.~\ref{fig:image}(b) shows the image of Alcyone within the aperture. The central columns are saturated. Additionally, pixels in column 668 are saturated by the bleed column of the nearby star 24~Tau. We construct each light curve from the unsaturated pixels in the halo. Three-day segments of the light curves of the four pixels highlighted in Fig.~\ref{fig:image}(b) are shown in Fig.~\ref{fig:pixlc}. The light curve of each pixel clearly shows the change in flux caused by the pointing drift, with corrections made by thruster firings at $\sim$6\,h intervals. Further details of K2 data characteristics are provided by \citet{vancleve16}.

The simple aperture photometry light curve using all the pixels within the target pixel mask is shown in Fig.~\ref{fig:methcomp}(a). A large oscillatory signal correlated with the pointing drift is apparent in this light curve, and is caused by aperture losses as Alcyone moves across the mask.

Several pipelines have been developed to process K2 data and remove these pointing-drift systematics, such as \textsc{k2sff} \citep{vanderburg14}, \textsc{k2p2} \citep{lund15}, \textsc{k2sc} \citep{aigrain15,k2sc}, \textsc{k2varcat} \citep{armstrong15}, \textsc{k2phot} \citep{vaneylen16}, and \textsc{everest} \citep{everest,everest2}. While these methods have been highly-successful for fainter stars, even approaching the photometric precision of the nominal \textit{Kepler} mission, they have substantial difficulties when dealing with the large aperture losses that arise with these target pixel masks. Fig.~\ref{fig:methcomp}(b) shows a segment of the Alcyone light curve after processing with the \textsc{k2sc} pipeline. Although the signal due to the pointing drift has been substantially reduced in this light curve, it has not been completely removed and still dominates the stellar signal. An alternative photometric method is required.

\subsection{Method}
Each circular aperture is dominated by the light of the bright central star it contains. The first major feature is saturation, where each of the core pixels bleeds flux in both directions along its column. Even for very saturated targets, this process is conservative, with the consequence that extremely elongated apertures have been used to perform photometry at the cost of many pixels \citep{kolenberg11}. Other than this, there is a complex, position-dependent point spread function (PSF), resulting from both the usual diffraction through the limited telescope aperture and from multiple orders of reflection in the telescope optics, which makes a significant contribution for these bright stars. To first order, this resembles an image of the telescope pupil; higher orders of reflection impose structure related to the focal plane CCD array, and multiple distorted images of both the array and the pupil. It is therefore not straightforward to model the PSF to perform photometry \citep{bryson10}, as is common in other contexts \citep{schechter93} and has been previously done with \textit{Kepler} \citep{libralato16}. Instead, we adopt a different approach to optimally extract a light curve from this complex `halo' of scattered light.

\begin{figure}
	\includegraphics[width=\columnwidth]{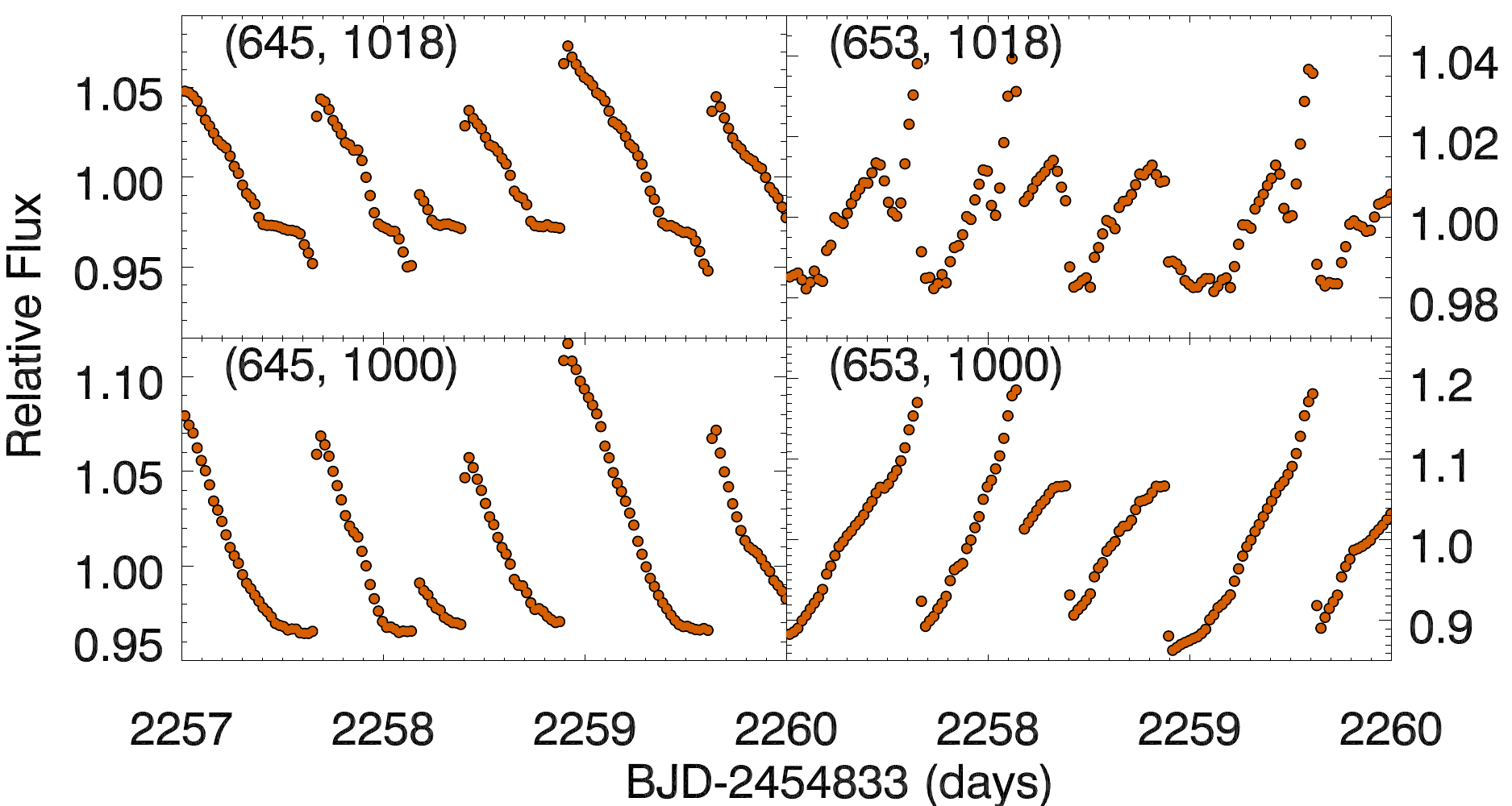}
	\caption{Three day segments of the four pixels highlighted in Fig.~\ref{fig:image}(b). The pixel coordinates are labelled. The drift signal clearly dominates, but is anti-correlated on either side of the roll motion, which is primarily from left-to-right in this figure.}
	\label{fig:pixlc}
\end{figure}

As seen in Fig.~\ref{fig:pixlc}, the pointing drift signal is anti-correlated on either side of the roll motion, while the stellar signal will be correlated across the aperture. This means we can separate the spatially-dependent pointing drift signal from the stellar signal and other instrumental signals that may be correlated across the aperture, such as focus drift. On this basis, we propose a new method to construct a light curve from a weighted sum of the non-saturated pixels, with the weights chosen to minimize pointing systematics and aperture losses. This is therefore a simultaneous method for aperture photometry and systematics correction, which are ordinarily done in separate stages. We refer to this method as halo photometry.

% Figure 2

The flux $f_i$ of the final light curve at each observation $i$ is chosen to be
\begin{equation}
	f_i = \sum_{j=1}^{M} w_j p_{ij},
\end{equation}
\noindent where $w_j$ is the weight of pixel $j$, $p_{ij}$ is the flux in pixel $j$ at observation $i$, and $M$ is the number of pixels. The weights $w_j$ are all defined to be positive and their sum is constrained to be unity. If the weights were all constant, this would reduce to simple aperture photometry. Instead, we have a soft aperture, and a suitable objective function must then be defined in order to find the optimal weights, of which several choices are possible. While a band-limited PSF would allow for negative weights in principle, we constrain these to be strictly positive in order to avoid signal self-subtraction.

\begin{figure}
	\includegraphics[width=\columnwidth]{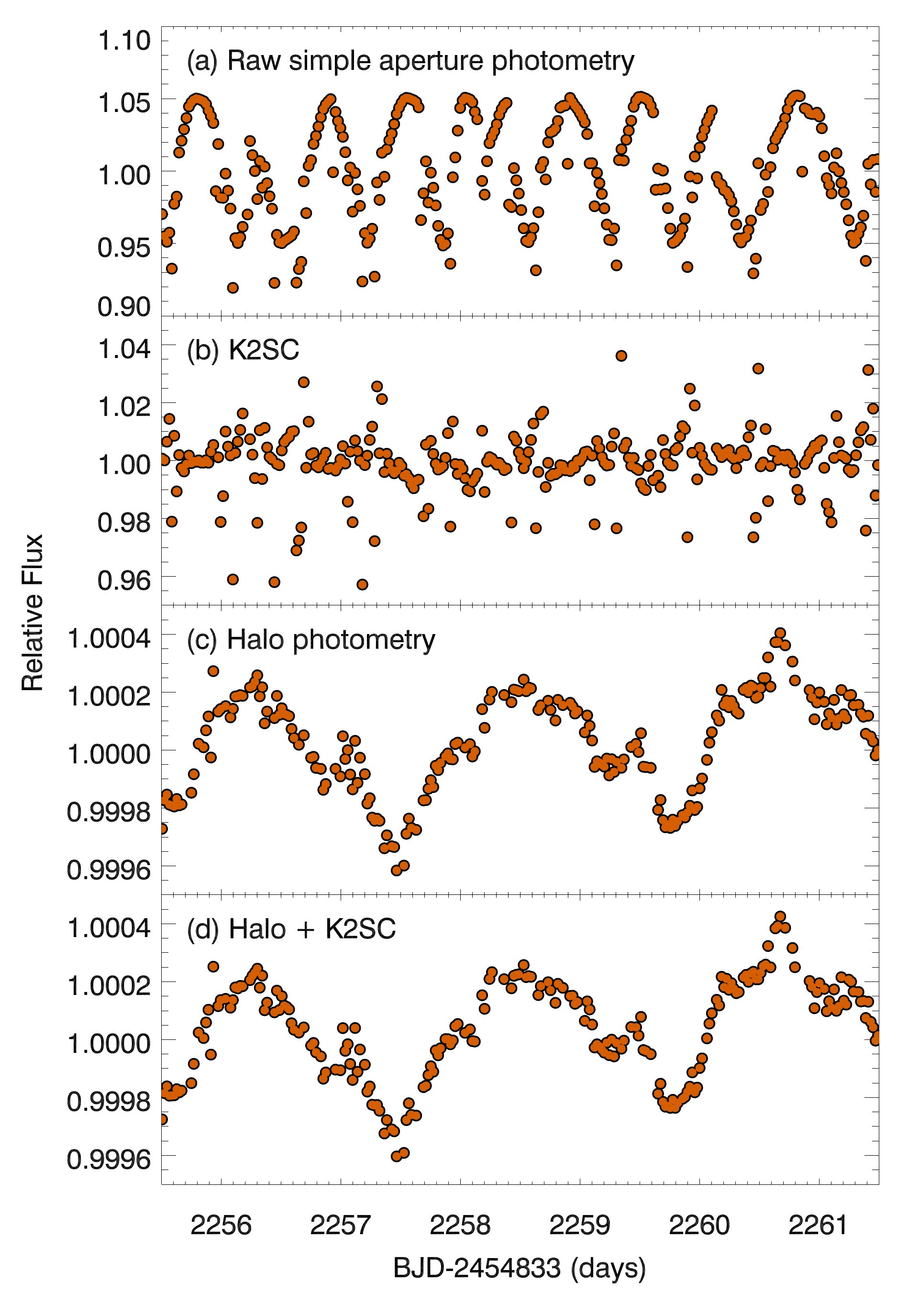}
	\caption{Six day segment of the Alcyone light curve processed with different methods: (a) raw simple aperture photometry, (b) the \textsc{k2sc} pipeline, (c) halo photometry, and (d) halo photometry with post-processing with \textsc{k2sc}. Note the change in scale between panels.}
	\label{fig:methcomp}
\end{figure}

Noting that the jumps caused by the thruster firings in K2 are large and sudden, whereas stellar variability is generally more gradual, we are inspired to consider minimizing the differences between consecutive observations across the time series, that is, we minimize the normalized first-order total variation (TV) of fluxes given by
\begin{equation}
\text{TV} = \dfrac{\sum_{i=1}^{N} |{f}_i - {f}_{i-1}|}{\sum_{i=1}^{N} {f}_i}, \label{eqn:tv}
\end{equation}
\noindent where $N$ is the total number of observations. Because the jumps occur in opposite directions on either side of the roll motion, and the wide diversity in pixels can compensate for inter- and intra-pixel sensitivity variation, the right combination of weights can effectively remove jumps. Furthermore, because stellar variability is present in all pixels and the weights are constrained to be positive, it is hard to spuriously suppress real stellar variability, even in cases where this variability is sudden such as a flare or transit event. We also consider second-order TV, where centred second finite differences replace first differences as above. Higher orders can likewise be defined, but in practice, we find that first or second order differences are sufficient for our present purpose. We note that TV need not be normalized by the mean flux term we include in the denominator, but we introduce this normalization to avoid finding the trivial solution where dark pixels are chosen to minimize total flux and therefore spuriously minimize absolute TV. We apply a sequential least squares programming algorithm implemented in \textsc{SciPy} \citep{jones_scipy_2001} to find the weights that minimize equation~(\ref{eqn:tv}), and the algorithm converges quickly on a modern laptop. 

Fig.~\ref{fig:image}(c) shows the final weights for the light curve of Alcyone, with up-weighted pixels shown by red shades, while down-weighted pixels are indicated by blue shades. Comparison with the image in Fig.~\ref{fig:image}(b) reveals that the diagonal halo features, consisting of the diffraction spikes and the higher-order reflection artefacts of the telescope spiders, are heavily up-weighted. A segment of the resulting light curve is shown in Fig.~\ref{fig:methcomp}(c). The pointing drift signal has been successfully removed, leaving behind the stellar variability.

An alternative objective function to TV would be to use the normalized quadratic variation (QV), that is, the sum of the square of differences between consecutive observations,
\begin{equation} 
\text{QV} = \dfrac{\sum_{i=1}^{N} \left({f}_i - {f}_{i-1}\right)^2}{\sum_{i=1}^{N} {f}_i}. \label{eqn:ls}
\end{equation}
We have investigated the use of QV for this application, and found it achieved visibly poorer results on real and simulated data compared to TV. 

To understand why TV may be preferable to QV, it is worth noting that TV has been very widely applied as a regularizing term in many domains of signal processing, especially for image data and for de-noising step-wise constant functions with high frequency noise \citep{rudin92,strong03}. These signal processing applications have their foundation in compressed sensing theory. For a continuous function, the TV is the arc length between two points; in the discrete case, this becomes the sum of absolute values of finite differences. In the language of Minkowski metrics in linear algebra, the sum of absolute values of a vector's elements is called the $\ell1$ norm, also known as the taxicab or Manhattan metric: this is the distance between two points subject to the restriction that you can only move parallel to the axes, as on a grid of streets \citep{menger52}. Other metrics include the square root of the sum of squares of the components of a vector, which is the familiar $\ell2$ or Euclidean norm, while the number of nonzero elements of a vector in some basis, a measure of sparsity, is called the $\ell0$ norm. So TV is the $\ell1$ norm on the derivative of a function, whereas QV is the $\ell2$ norm. From compressed sensing theory, it can be shown that by minimizing $\ell1$ norms under certain general conditions (which is computationally easy) it is also possible to minimize the $\ell0$ norm (which is otherwise computationally hard), and therefore that TV is the appropriate objective function for enforcing sparsity in the gradient of a time series or image \citep{candes04}. Therefore, if we assume that a true signal has a small gradient with respect to time nearly everywhere, it is often effective to minimize TV to constrain a reconstruction from noisy data. We do not see a clear reason why the gradient of stellar variability should be sparse, and employ the method \emph{ad hoc}, rather than explicitly as a case of compressed sensing. 

Pixel-level decorrelation (PLD) in K2 has been explored by \citet{everest}, extending earlier work by \citet{demingpld}, using least-squares ($\ell2$) methods to project light curve residuals onto principal components of normalized pixel time series. In their approach, simple aperture photometry is first used to obtain a raw light curve, which is then used to normalize the raw pixel time series. In \citet{everest}, the further step is taken of generating second- or third-order polynomials in these time series, to obtain a large basis set spanning nonlinear components of the aperture losses. We note that $\ell2$ approaches to weighted-pixel soft aperture photometry were proposed by \citet{jenkinsweights}, and independently investigated by Hogg, Foreman-Mackey \& Goodman (unpublished), but we do not know of any previous TV-based method.

\begin{figure}
	\includegraphics[width=1\columnwidth]{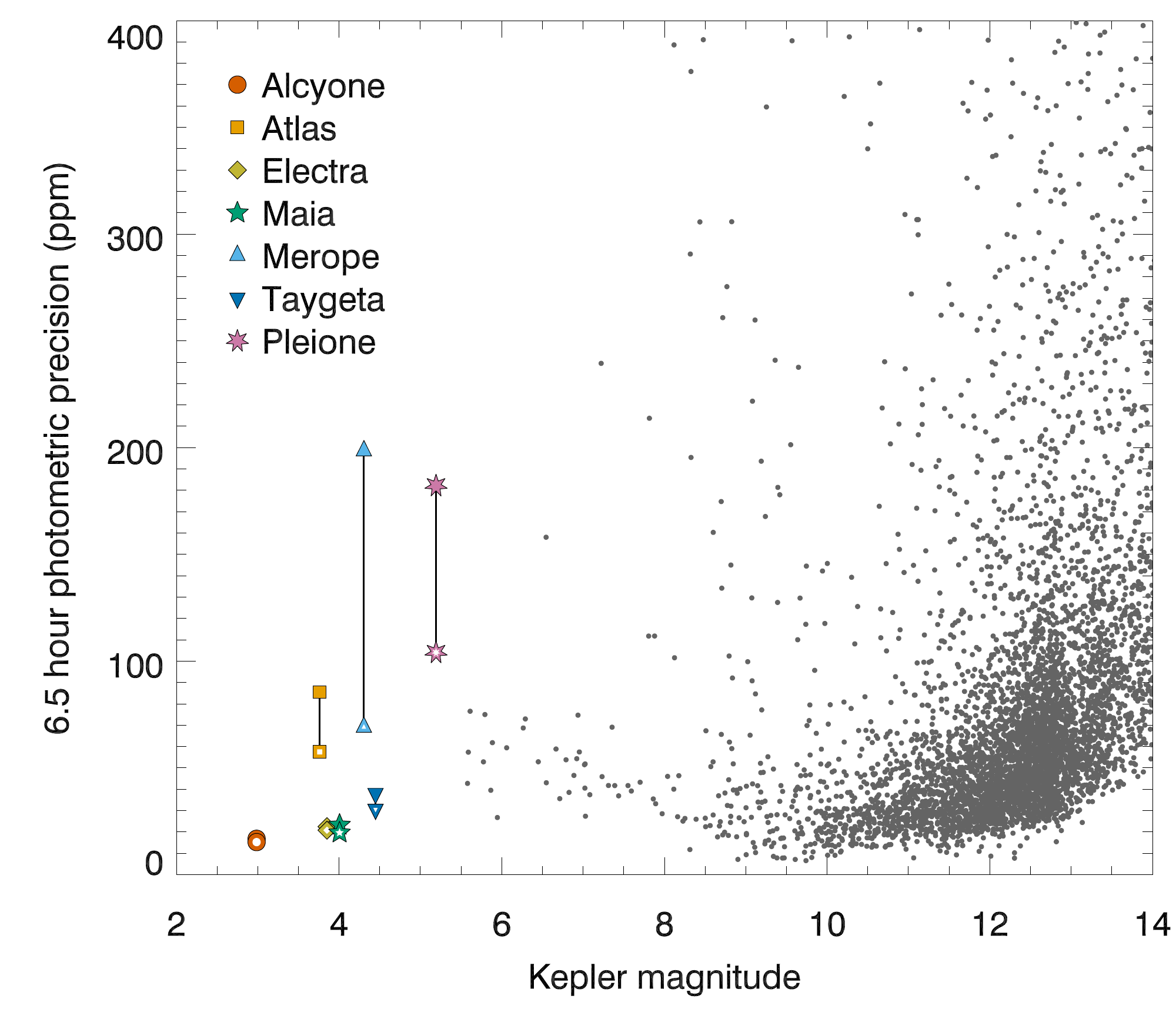}
	\caption{Photometric precision as a function of \textit{Kepler} magnitude. The solid symbols indicate light curves prepared with the halo method alone, while open symbols are for those with further corrections using \textsc{k2sc}; these two measurements are connected by solid lines for each star. For comparison, the grey dots indicate CDPP for K2 C4 dwarfs ($\log g > 4$) in the Ecliptic Plane Input Catalog \citep{huber16} using \textsc{k2sc}. }
	\label{fig:cdppmag}
\end{figure}

Halo light curves obtained as above typically contain small residual roll systematics and more substantial long-term drifts. We correct residual roll systematics using \textsc{k2sc}, a Gaussian Process-based K2 detrending pipeline which jointly models stellar variations and instrumental systematics \citep{k2sc}. We use as the $x,y$ position inputs the corrected inputs of the nearest standard light curve available in MAST. A segment of the Alcyone light curve after \textsc{k2sc} post-processing is shown in Fig.~\ref{fig:methcomp}(d). For the longer-term systematic drifts, we fit and subtract a polynomial of variable order, typically eight or nine, to remove these trends. The trend removal affects frequencies lower than 0.2\,d$^{-1}$. The lightcurve of Maia exhibits a long-period signal that is intrinsic to the star, as well as the longer-term systematic drifts. To avoid overfitting in this case, we use the polynomial fit of nearby Taygeta to set the high-order coefficients, while allowing coefficients up to the quartic term to vary. While it would be preferable to fit a linear combination of cotrending basis vectors \citep[CBVs;][]{smi+12,stu+12}, those available for K2 unfortunately are dominated by thruster firing systematics and are unsuitable to fit to otherwise-clean light curves possessing only long-time-scale trends. Furthermore, the long-time-scale common-mode systematics in halo photometry are not yet well-quantified, and it is not clear whether the CBV approach is well-suited to these light curves. We leave this as the subject for future work.

With the large diversity of available pixels there is the potential for over-fitting. Spurious signals might be introduced, or real astrophysical variability removed, either by the TV objective function suppressing time variability, or by inappropriately weighting noisy pixels. To account for the former, it would be ideal to divide up the light curve into separate chunks, and use these separately for training and validation. For instance, one might train weights on the first half of the light curve, and use these to extract photometry from the second, and vice versa. This is a computationally-expensive approach that we have not adopted.

We considered it important to address the problem of pixel-level abnormalities by checking whether the same light curve could be reproduced from independent subsets of pixels. To select independent sets of pixels with similar overall distribution, we unravel the array coordinates, and select eight ensembles taking every eighth pixel time series. We then applied the method on these subsampled pixel arrays and produced light curves for each. We find that the long-term trend behaviour of all of these light curves differs, but that after filtering, subsampled light curves reproduce the same overall behaviour as the globally-optimized light curve. A consensus light curve, consisting of the means of the subsampled light curves, is very similar to the globally-optimized light curve, again except for an overall trend. Finding no reason to prefer one or the other, we use globally-optimized light curves for the scientific analysis in the remainder of this paper. 

\begin{figure}
	\includegraphics[width=1\columnwidth]{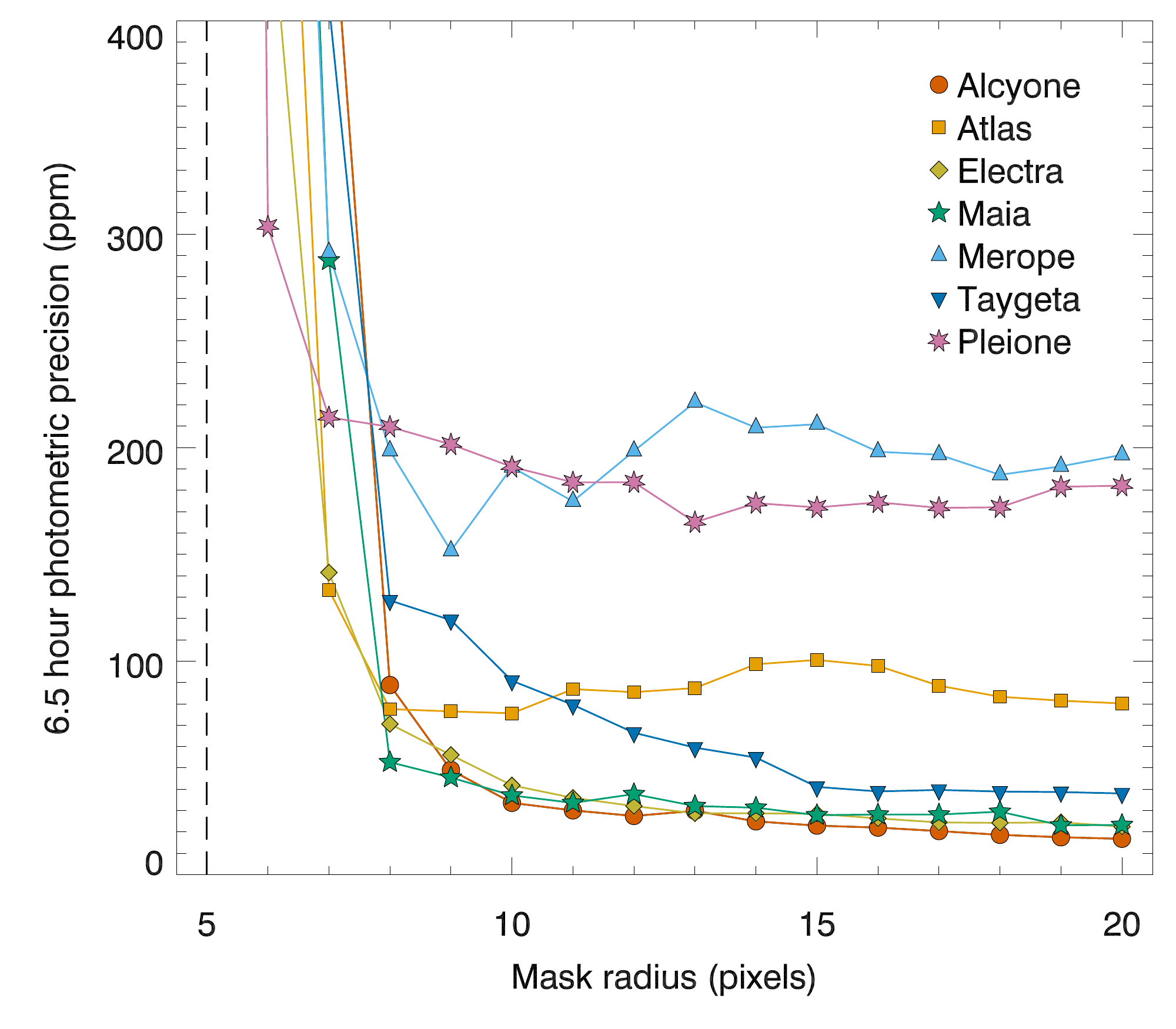}
	\caption{Photometric precision as a function of mask radius for each star. The dashed line at a mask radius of five pixels indicates the region below which almost all pixels are saturated.}
	\label{fig:masksize}
\end{figure}

%\section{Results}
\section{Photometric Precision} \label{precision}
The efficacy of K2 data processing pipelines has commonly been determined by measuring the 6.5\,h Combined Differential Photometric Precision \citep[CDPP;][]{jenkins10,cdppdef}. This measurement, however, is strongly influenced by the presence of stellar variability in a time series, so a straight application of CDPP is not informative in such a situation. When benchmarking K2 pipelines, CDPP values are therefore usually reported for dwarf stars for which stellar variability is minimal.

We do not have any bright and photometrically-quiet dwarf stars to test the halo photometry method on, so an alternative approach is required. We clean the light curves of the stellar signal by iterative sine-wave fitting (also called prewhitening) using the program \textsc{Period04} \citep{lenz05}, and then measure the CDPP of the residual light curves. We use the \citet{k2sc} CDPP-equivalent, which first removes long-term trends with a Savitzky-Golay filter, then takes the standard deviation of the means of all consecutive 13-sample (6.5~h) segments of the light curve, ignoring outliers more than $5\sigma$ away from the mean. 

The CDPP for each star is shown in Fig.~\ref{fig:cdppmag}, both for light curves processed using our halo photometry method alone (solid symbols) and with further processing with the \textsc{k2sc} pipeline to remove any remaining systematics results (open symbols). We find that \textsc{k2sc} typically reduces the CDPP of halo light curves by $\sim 10$ per cent, although in the cases of Atlas, Merope, and Pleione, the CDPP is reduced by \textsc{k2sc} by 32, 65,~and~42 per cent, respectively.

\begin{figure*}
	\includegraphics[width=1\columnwidth]{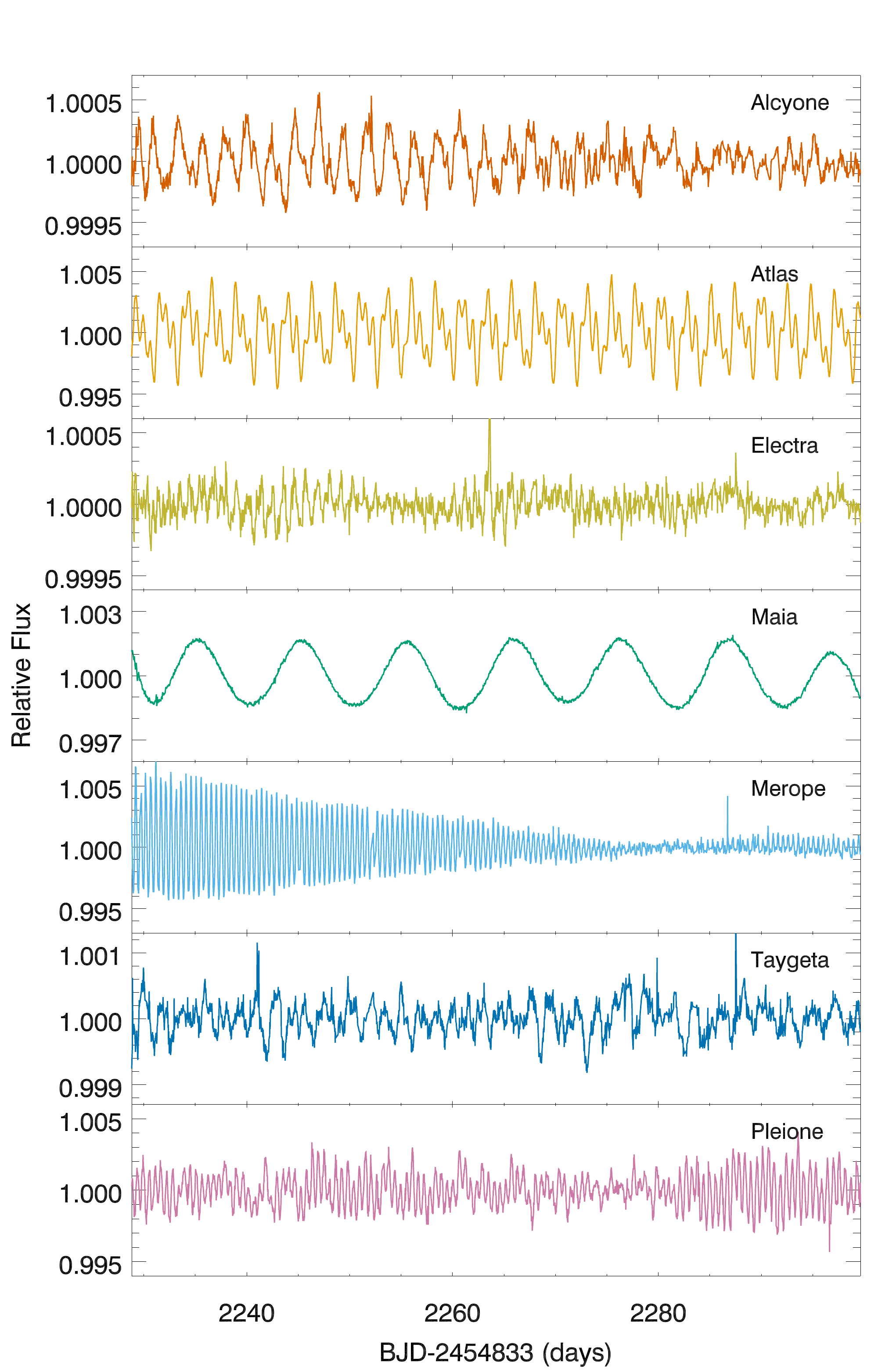}
	\includegraphics[width=1\columnwidth]{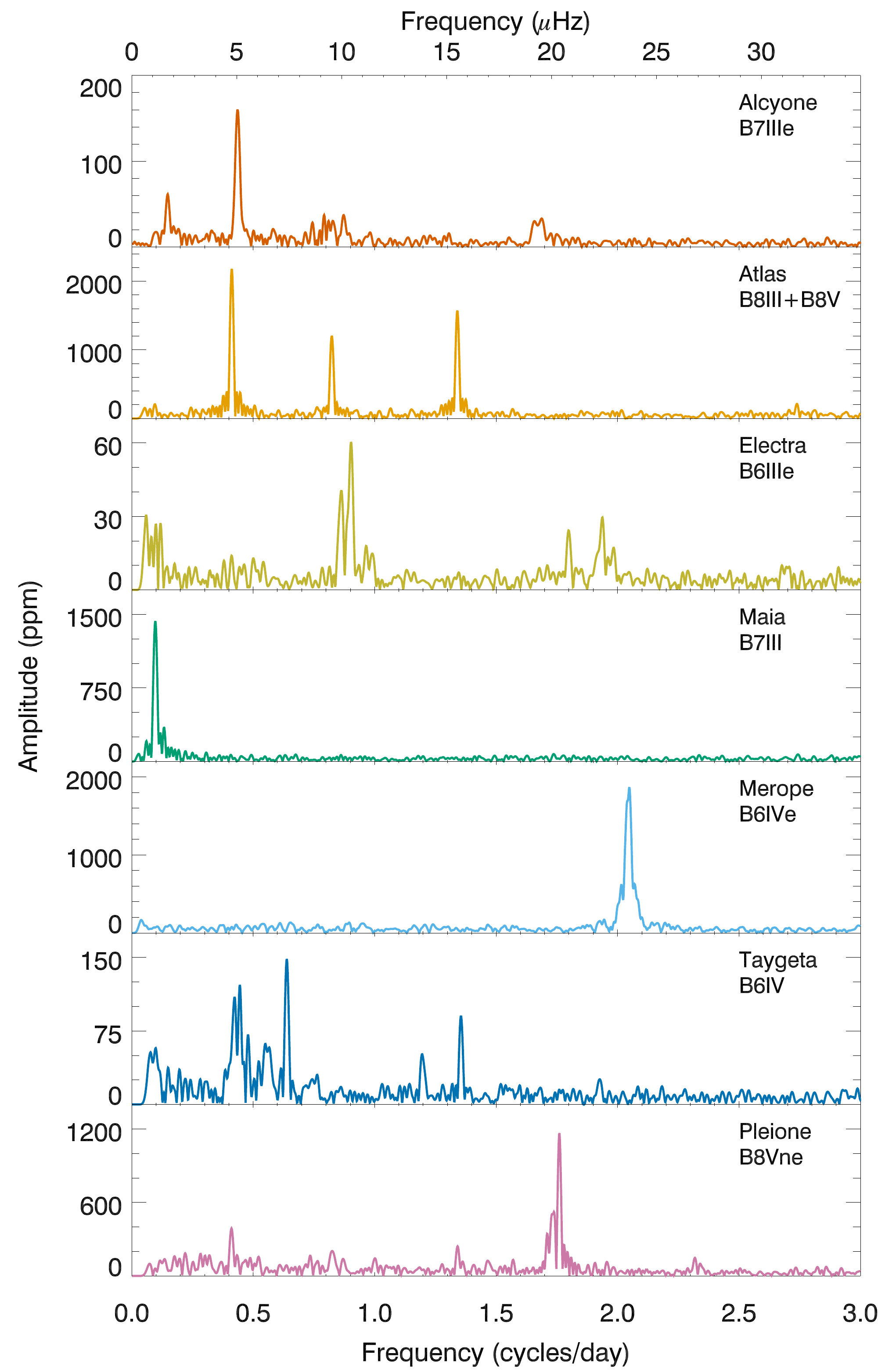}
	\caption{Time series (left) and amplitude spectra (right) of the brightest stars in the Pleiades. Brightest to faintest objects are listed top to bottom. Note the changes in scale on the y-axes.}
	\label{fig:res}
\end{figure*}

Comparison of the CDPP of the halo targets with the CDPP of fainter dwarf stars processed with the \textsc{k2sc} pipeline (grey dots in Fig.~\ref{fig:cdppmag}) shows that the precision from halo photometry is generally as good as that obtained for regular targets. For three of the halo targets, however, the CDPP is significantly higher than expected. We note that these three stars -- Atlas, Merope and Pleione -- have the highest amplitude stellar variability (see Section~\ref{pleiades}), and are significantly improved by further processing using \textsc{k2sc}. We attribute the increased CDPP to the halo photometry algorithm being less effective at suppressing jumps in the light curve in the presence of high-amplitude stellar variability. 

The fact that the CDPP is comparable to that of fainter stars, and even as good in some cases as the best CDPPs of the fainter stars, is indicative that this method delivers photometry of the quality required in principle to detect exoplanet transits. It is nevertheless important to establish in future work whether the TV minimization removes planetary transit signals as sharp, short-duration events: we suspect that because TV preserves $\sim$~hour-scale stellar oscillations, as shown in Section~\ref{pleiades}, that this is not likely to be a problem for transit events on this time-scale. This future work will involve transit injection tests to establish the exact degree to which transit signals are dampened due to overfitting, and whether the signal could be preserved by excluding known transits from the TV optimization step.

As halo photometry is intended as a solution to the problem of pixel allocation for very bright objects, it is important to determine how its precision varies with the size of the allocated pixel mask. Furthermore, it may be important to avoid contamination from other nearby bright sources to reduce the mask size beyond the requirements imposed by bandwidth constraints. The raw K2 data were obtained with 20~pixel radius circular masks; we run the algorithm separately on masks with radii from 6~to the full 20~pixels and measure the 6.5~h CDPP as a function of mask radius. At 5~pixels and below, almost all pixels are saturated. The results are shown for all seven stars in Fig.~\ref{fig:masksize}. The precision follows an L shaped curve: radii much less than 8~or 9~pixels give very poor photometric precision, and there is some improvement, albeit limited, for radii much larger than this. The bump around $\sim15$~pixels for Atlas corresponds to the position of the nearby background star 2MASS J03491192+2403507, and we interpret this as the result of contamination. We conclude that a mask with a 12~pixel radius is generally sufficient to achieve good precision, which is equivalent to approximately two to four $Kp$\,=\,12\,mag stars.

\begin{figure*}
	\includegraphics[width=2\columnwidth]{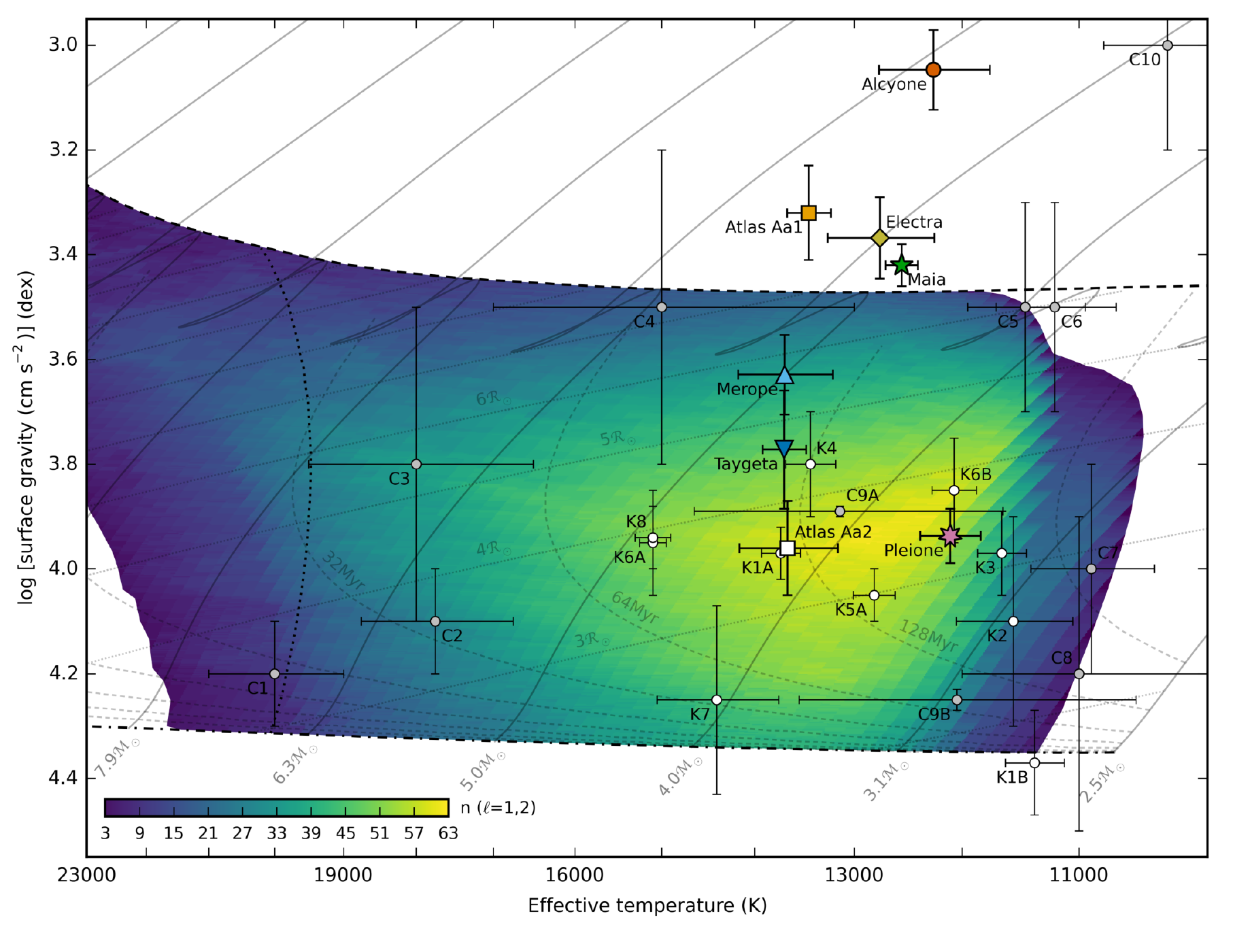}
	\caption{Kiel diagram ($\log g$ -- $T_\mathrm{eff}$ diagram) of the brightest Pleiades stars and the sample of B-type stars for which in-depth seismic analysis has been made possible by the CoRoT (grey circles) and \textit{Kepler} (white circles) missions. Colours and symbols for the Pleiades stars are the same as in Fig.~\ref{fig:cdppmag}, with the addition that the B8V secondary component of Atlas is indicated by the white square to distinguish it from the B8III primary (orange square). The following annotations are used for the CoRoT and \textit{Kepler} B stars: C1 -- HD\,48977, C2 -- HD\,43317, C3 -- HD\,50230\,A, C4 -- HD\,50846\,A, C5 -- HD\,182198, C6 -- HD 181440, C7 -- HD\,46179, C8 -- HD\,174648, C9 -- HD\,174884\,AB, C10 -- HD\,170935, K1 -- KIC\,4931738\,AB, K2 -- KIC\,10526294, K3 -- KIC\,7760680, K4 -- KIC\,3459297, K5 -- KIC\,6352430\,A, K6 -- KIC\,4930889\,AB, K7 -- KIC\,9020774, K8 -- KIC 11971405. The dot-dashed black line indicates the zero-age main sequence (ZAMS), while the dashed black line indicated the terminal-age main sequence (TAMS). The cool edge of the $\beta$\,Cep instability strip for radial ($\ell = 0$) modes is plotted with a dotted black line. Solid grey lines show evolutionary tracks for the masses indicated, dashed grey lines show isochrones for ages from 2$^0$ to 2$^8$\,Myr, and dotted grey lines show isoradii for integer multiples of the solar radius. The SPB instability strip is defined as the region for which at least three gravity modes of degree $\ell$ = 1 or 2 are excited, and is shown in colours ranging from purple (3 excited modes) to yellow (63 excited modes), as indicated by the colour bar. All model data are taken from \citet{moravveji16b} for $Z = 0.014$, the solar mixture of \citet{asplund09}, an exponential core-overshooting $f_\mathrm{ov} = 0.02$, and opacity enhancement factors for iron and nickel of $\beta_\mathrm{Fe} = \beta_\mathrm{Ni} = 1.75$.
	}
	\label{fig:kiel}
\end{figure*}

\section{Variability in the Pleiades} \label{pleiades}
The naked-eye Pleiades were the first collection of stars to be recognized as a cluster \citep{michell1767}, and together with many fainter members they remain one of the most prominent and well-studied open clusters. While the cluster contains over 1000 members \citep[e.g.][]{bouy15}, with an age of $\sim$125\,Myr \citep{stauffer98} it is visually dominated by several hot, massive stars. The seven brightest stars in the Pleiades, which we study here -- Alcyone, Atlas, Electra, Maia, Merope, Taygeta, and Pleione -- are all late B-type stars. Their time series and amplitude spectra are shown in Fig.~\ref{fig:res}. We search all seven light curves using the \textsc{k2ps} planet-search package \citep{k2ps,2016mnras.461.3399p}, but do not find any evidence of transiting planet candidates or eclipses.

It is remarkable to note that, although these stars are from the same cluster, and therefore have a similar age and initial composition, and they span a narrow range in mass and effective temperature, they display a diverse range of variability, both in terms of amplitude and frequency content. Such diversity would not be able to be seen from the ground; periods in the range 0.5--3\,d are difficult enough, but amplitudes below a mmag are impossible from single-site data \citep[see e.g.][]{decat02}. It has only been with the dawn of space-based photometric missions, such as MOST, CoRoT and \textit{Kepler} that such low amplitude variability could be detected \citep[e.g.][]{aerts06,degroote09,balona11}.

Variability may arise from a number of phenomena, including various pulsation mechanisms and rotational modulation. These types of variability may appear similar, and further knowledge of the star's properties can be useful to make the distinction. Properties of each star are provided in Table~\ref{tab:prop}. 

The locations of the stars in the Kiel diagram ($\log g$ -- $T_\mathrm{eff}$ diagram)  are shown in Fig.~\ref{fig:kiel}, alongside other B-type stars that have been studied in-depth using CoRoT and \textit{Kepler} data, and the theoretical slowly pulsating B (SPB) star instability strip calculated by \citet{moravveji16b}. The models used to determine the instability strip exclude the effects of rotation, however most of the hot Pleiades stars are rapid rotators. Models have indicated that rapid rotation expands the region over which SPB pulsations may be excited \citep[e.g.][]{salmon14}. The CoRoT target HD\,170935 (marked as C10 in Fig.~\ref{fig:kiel}) shows SPB pulsations despite its more-evolved state \citep{degroote11}. Therefore, each of the hot Pleiades stars may potentially be a SPB star.

\begin{table*}
	\centering
	\caption{Properties of the bright Pleiades}
	\label{tab:prop}
	\begin{tabular}{lccccccccc} % four columns, alignment for each
		\hline
		Name      & HD    & Sp. type & $V\,(\mathrm{mag})$ & $T_\mathrm{eff}\,(\mathrm{K})$ & $\log (g / \mathrm{cm\,s^{-2}})$ & $v\sin i\, (\mathrm{km\,s^{-1}})$ & $E(B-V)\,(\mathrm{mag})$ & $M\,(\mathrm{M_\odot})$  & $R\,(\mathrm{R_\odot})$ \\
		\hline
		Alcyone   & 23630 & B7IIIe & 2.87 & 12258$\pm$505$^a$ & 3.047$\pm$0.076$^a$ & 140$\pm$10$^a$ & 0.01$\pm$0.01 &      5.9$^b$      &  9.3$\pm$0.7$^c$  \\
		Atlas Aa1 & 23850 & B8III  & 3.84 & 13446$\pm$218$^d$ &  3.32$\pm$0.09$^c$  & 240$^e$        & 0.07$\pm$0.01 & 4.74$\pm$0.25$^e$ &  7.9$\pm$0.8$^c$  \\
		Atlas Aa2 &  ---  & B8V    & 5.52 & 13660             &  3.96$\pm$0.09$^c$  & 60$^e$         & 0.07$\pm$0.01 & 3.42$\pm$0.25$^e$ &  3.2$\pm$0.3$^c$  \\
		Electra   & 23302 & B6IIIe & 3.70 & 12754$\pm$504$^a$ & 3.368$\pm$0.078$^a$ & 170$\pm$12$^a$ & 0.00$\pm$0.01 &      4.7$^b$      &  6.3$\pm$0.7$^c$  \\
		Maia      & 23408 & B7III  & 3.87 & 12550$\pm$150$^c$ &  3.42$\pm$0.04$^c$ &  33$\pm$5$^f$  & 0.02$\pm$0.04 & 4.22$\pm$0.18$^g$ & 6.61$\pm$0.11$^c$ \\
		Merope    & 23480 & B6IVe  & 4.18 & 13691$\pm$481$^a$ & 3.629$\pm$0.076$^a$ & 240$\pm$14$^a$ & 0.06$\pm$0.02 & 4.25$\pm$0.08$^h$      & 4.79$\pm$0.17$^c$ \\ 
		Taygeta   & 23338 & B6IV   & 4.30 & 13696$\pm$222$^d$ & 3.772$\pm$0.113$^d$ & 105$\pm$16$^i$ & 0.02$\pm$0.01 & 4.41$\pm$0.09$^h$ & 4.36$\pm$0.14$^c$ \\
		Pleione   & 23862 & B8Vne  & 5.19 & 12106$\pm$272$^a$ & 3.937$\pm$0.052$^a$ & 286$\pm$16$^a$ & 0.05$\pm$0.02 &      3.8$^b$      & 4.17$\pm$0.17$^c$ \\
		\hline
	\end{tabular}
    References: (a) \citet{fremat05}, (b) \citet{zorec05}, (c) derived here, (d) \citet{david15}, (e) \citet{zwahlen04}, (f) \citet{royer02}, (g) \citet{kochukhov06}, (h) \citet{zorec12}, (i) \citet{abt02}.
\end{table*}

Four of the stars are known Be stars (Alcyone, Electra, Merope, and Pleione), meaning they have shown emission lines in their spectra associated with a circumstellar `decretion' disc \citep[for a recent review of Be stars, see][]{rivinius13}. Pleione is known to transition between a Be phase and a shell phase -- the difference between such phases is understood to be the orientation of the disc, with shells occurring when it obscures the central star \citep{struve1931}. Spectra contemporaneous with the K2 observations available at the Be Star Spectra (BeSS) database\footnote{\url{http://basebe.obspm.fr}} \citep{neiner11} show all four stars presently have emission features, and Pleione is in a shell phase. These circumstellar discs can also be a potential source of variability in the time series. 
Many Be stars are known to pulsate \citep{baade82,bolton82,rivinius03,walker05,saio07,neiner09}; \citet{aerts06b} concluded that the pulsating Be stars are not a separate class of variable stars, but rapidly-rotating and complex examples of the SPB stars. There is evidence to suggest that these pulsations are responsible for mass-loss episodes that launch stellar material into the decretion disc \citep{rivinius98,rivinius03,huat09,baade16}, and a model has been developed to explain how pulsations, coupled to rapid, sub-critical rotation can achieve this \citep{kee14}.

A large fraction of B stars are in multiple systems; mid-B stars have, on average, a total multiplicity frequency of 1.3$\pm$0.2 \citep{moe16}. It is convenient that the Pleiades, lying near the ecliptic, are amenable to lunar occultation measurements capable of resolving close binaries. Three of the hot Pleiades are known to be in multiple systems -- Atlas, Taygeta, and Pleione. Of these, the secondary companion of Atlas is also a B star, and its properties are provided alongside the others in Table~\ref{tab:prop}. Atlas Aa2, as it is referred to in the Washington Double Star Catalog \citep[WDS;][]{mason01}, may also be a source of variability in the light curve. Taygeta and Pleione have faint, low-mass companions.

The final cause of variability we need to consider is rotational modulation caused by the presence of surface features. For rotation to be a plausible explanation, two conditions need to be met by the implied rotational velocity at the equator; the velocity must be at least as large as the observed projected rotational velocity, $v \sin i$, and it must be smaller than the critical velocity. To test these conditions, the radius of the star must be known, as well as  $v \sin i$ and the potential rotation period. 

We have been conducting an observing campaign to measure the angular diameters of these stars using the Center for High Angular Resolution Astronomy (CHARA) Array. We present the results of this campaign for Maia in Section~\ref{sect:maiainterf}. Interferometric results for the other stars will be presented elsewhere; additional observations from multiple position angles are required to adequately sample the shape of the rapidly rotating Pleiades stars. In the meantime, we estimate the radii of these stars using other methods. For Atlas, we first estimate the angular diameters using the $V-K$ surface brightness relation \citep{kervella04,boyajian14}. The system has $V\,=3.63\,\mathrm{mag}$ and $K\,=3.76\,\mathrm{mag}$, while \citet{pan04} find magnitude differences of $\Delta V = 1.68\pm0.07$\,mag and $\Delta K = 1.86\pm0.06$\,mag between the components from interferometry. We subsequently estimate Atlas~Aa1 to have a radius of $7.9\pm0.8\,\mathrm{R_\odot}$, and Atlas~Aa2 to have a radius of $3.2\pm0.3\,\mathrm{R_\odot}$, assuming a conservative 10 per cent uncertainty in the angular diameter and using Atlas' geometric parallax \citep[132$\pm$4 pc;][]{zwahlen04}. 

For the Be stars, the use of the $V-K$ relation is not appropriate because the circumstellar discs contribute significant flux, particularly in the infrared. \citet{touhami13} argued that it is better to compare model and observed fluxes in the ultraviolet where the disk contribution is negligible. We gathered the available International Ultraviolet Explorer (IUE) spectra from the NASA Mikulski Archive for Space Telescopes (MAST) at STScI\footnote{\url{http://archive.stsci.edu/iue/}}, and repeated the analysis of \citet{touhami13}, comparing the spectra with Kurucz model fluxes, in order to determine the angular diameter. We assume a binary companion for Taygeta with $T_\mathrm{eff}$\,=\,8306\,K, $\log g$\,=\,4.2\,dex, and $\Delta K$\,=\,2.5\,mag \citep{richichi94}, and a binary companion for Pleione with $T_\mathrm{eff}$\,=\,6650\,K, $\log g$\,=\,4.34\,dex, and $\Delta K$\,=\,2.1\,mag \citep{touhami13}. \citet{touhami13} also determined the angular diameters of Pleione and Alcyone; our result for Alcyone is in perfect agreement. For Pleione, we excluded spectra taken early in the IUE mission when Pleione was in a shell phase, during which FUV flux declined because of disc obscuration \citep{doazan93}. This left the 7 SWP (115--190\,nm) and 6 LWP (180--330\,nm) spectra from 1993--4. We subsequently find Pleione to have a larger UV flux and radius than found by \citet{touhami13}. We determine the linear radii using the VLBI parallax measurement for the Pleiades \citep{melis14}. The radii are listed in Table~\ref{tab:prop} alongside the other properties from the literature.

We note there are several limitations to these estimates of the radii. The method of \citet{touhami13} assumes spherical stars with no gravity darkening, so the resulting angular diameters will be intermediate between the actual polar and equatorial angular diameters. Furthermore, the observed $v \sin i$ is likely to be lower than the true value because gravity darkening results in the most rapidly rotating part of the star being less conspicuous \citep{townsend04}. We therefore only take these values of the radii and projected rotational velocities as indicative of the rotational time-scales for these stars.

In the remainder of this Section, we discuss the variability of each star in turn, in conjunction with observations recorded in the literature, to deduce the likely nature of each target.

\subsection{Alcyone}\label{sect:resalcyone}
Alcyone ($\eta$\,Tauri, 25\, Tauri, HR\,1165, HD\,23630) is a blue giant of spectral type B7IIIe, and is the brightest star in the Pleiades ($V$\,=\,2.87\,mag). Several visual companions are listed in the Washington Double Star Catalog \citep{mason01}, the closest of which is separated by 77 arcsec. While some lunar occultation measurements have suggested the existence of a closer companion with a separation between 1--30 mas and a magnitude difference of 1.6\,mag \citep{bartholdi75b,deVegt76}, others have been consistent with a single star scenario \citep{eitter74,mcGraw74,bartholdi75}. \citet{jarad89} identify a possible spectroscopic period of 4.13\,d. Alcyone appears single in speckle interferometry \citep{mason93,mason96,fu97} and adaptive optics observations \citep{roberts07}.

Being a Be star, Alcyone is girt by a circumstellar disc. \citet{quirrenbach97} and \citet{tycner05} have both measured the extent of the H\,$\alpha$ emission in the disc using long-baseline optical interferometry, finding the angular size of the semi-major axis to be 2.65$\pm$0.14 and 2.04$\pm$0.18 mas, respectively. Both found only small deviations from circular symmetry in the disc, suggesting the system is seen close to pole-on; \citet{quirrenbach97} found an axial ratio of 0.95$\pm$0.22, from which the minimum inclination angle is 18$^\circ$, while \citet{tycner05} found the axial ratio to be 0.75$\pm$0.05, implying a minimum inclination angle of 41$^\circ$. \citet{greznia13} found a much smaller angular extent of the disc in K band observations from the Palomar Testbed Interferometer, of only 0.71$\pm$0.01\,mas, although it must be noted that they fit only a single uniform disc component to their observations, and it seems likely the measurement was dominated by the central star. \citet{touhami13} were unable to resolve the disc in K band with the CHARA Array. None of these interferometric measurements found evidence of a companion.

The K2 time series and amplitude spectrum of Alcyone are shown in the top row of Fig.~\ref{fig:res}. The spectrum is dominated by a frequency at 0.4360$\pm$0.0004\,d$^{-1}$, although power is present at other frequencies as well, including at one-third of this primary frequency. There are low amplitude groupings of frequencies around 0.8~and~1.6\,d$^{-1}$. Such frequency groupings have been seen before in Be and regular B stars \citep{walker05,cameron08,diago09,huat09,neiner09,balona11,mcNamara12}. \citet{balona11} speculated that these groupings were caused by rotational modulation, however \citet{kurtz15}, using the excellent frequency resolution of the full \textit{Kepler} mission, were able to show that all peaks could be explained as simple combinations of a few modes. Such combination frequencies are indicative of non-linear coupling between pulsation modes. We therefore conclude that these low-amplitude peaks are caused by SPB oscillations. 

We measured the frequencies present in the time series by iterative sine-wave fitting; details are provided in Appendix~\ref{apx:freqs}, and the Fourier parameters of the peaks are given in Table~\ref{tab:Alcyone}. Which frequencies correspond to the uncombined, `base' frequencies cannot be reliably determined because any of the peaks can be described by a combination of others and \citet{kurtz15} showed that the base frequencies do not necessarily have the highest amplitudes. Furthermore, the frequency resolution of the K2 time series is quite limited, particularly in comparison to what was achieved with the full \textit{Kepler} mission. \textit{Kepler} observations show these frequency groupings contain many closely-spaced peaks, and it is quite likely that some of the peaks in the Alcyone amplitude spectrum are unresolved. Nevertheless, we are able to find possible decompositions of some modes into frequency combinations of other modes, which are indicated in Table~\ref{tab:Alcyone}, to illustrate that we are indeed seeing the frequency combination phenomenon.

The highest-amplitude frequency in the Alcyone amplitude spectrum may also be due to SPB pulsations. It occurs at a frequency close to half that of the frequency grouping at 0.8\,d$^{-1}$, and is somewhat broadened, suggesting that it is an unresolved peak of several frequencies. The peak at 0.15\,d$^{-1}$ can also be explained as a combination frequency.

Another possibility is that the highest-amplitude frequency arises from rotational modulation, or even from variability of the circumstellar disc. We can consider if this frequency occurs at an appropriate time-scale to be caused by rotation. We calculate the critical velocity from the mass and radius given in Table~\ref{tab:prop}, using this radius as an estimate of the equatorial radius. Rotation at the critical velocity would have a frequency of 0.61$\pm$0.07\,d$^{-1}$. The $v \sin i$ was found to be 140\,km\,s$^{-1}$ by \citet{abt02}; the minimum possible rotational frequency is 0.30\,d$^{-1}$. The peak at a frequency at 0.4362$\pm$0.0016\,d$^{-1}$ is therefore consistent with being caused by near-critical rotation.

\begin{figure}
	\includegraphics[width=1\columnwidth]{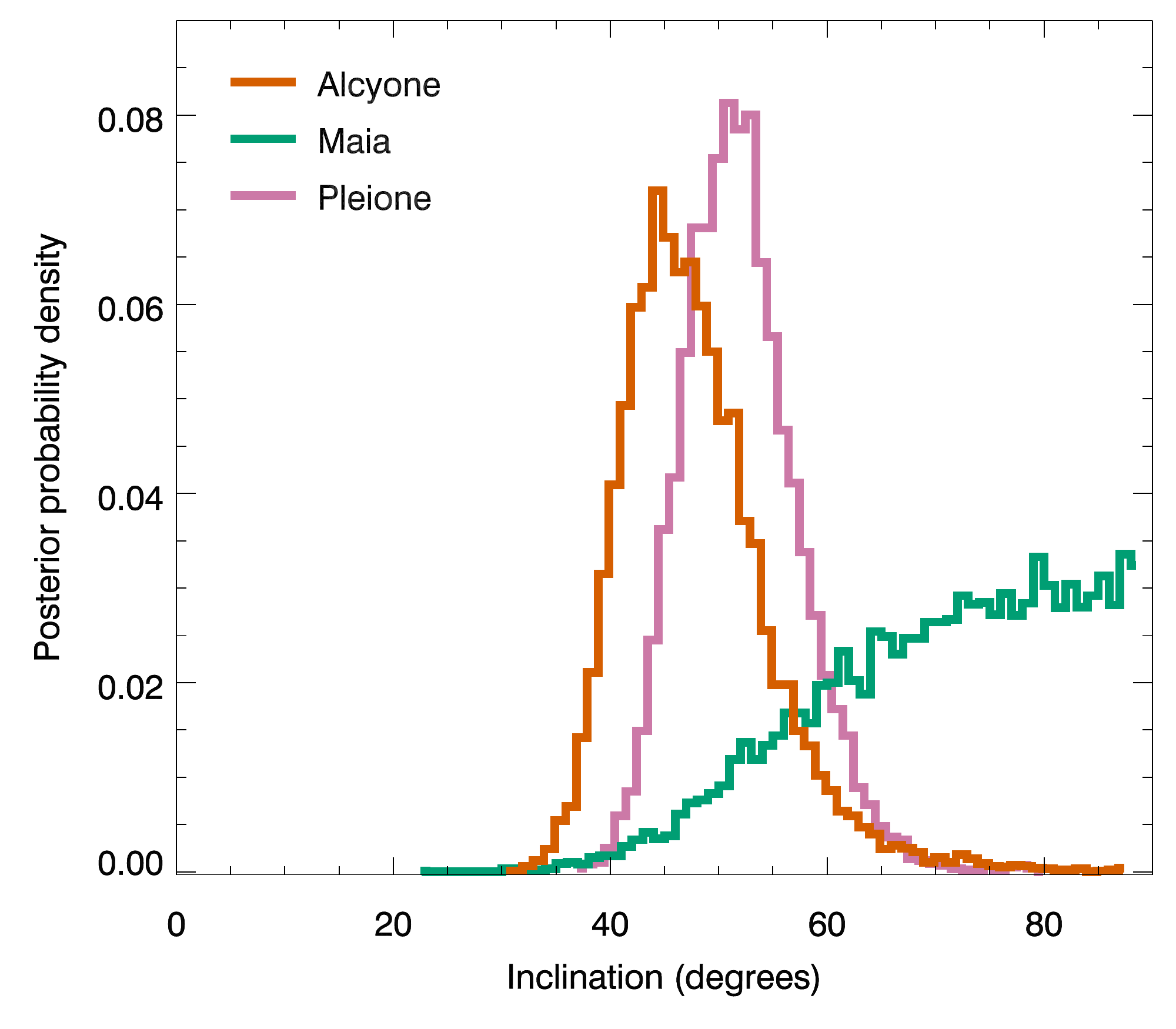}
	\caption{Posterior probability density functions for the possible inclinations of Alcyone, Maia, and Pleione calculated from their radii, projected rotational velocity, and possible rotation periods.}
	\label{fig:rotations}
\end{figure}

If this is the rotational frequency, then with an estimate of the stellar radii and the known $v \sin i$, the inclination of the star can be estimated. In order to account properly for the prior distribution of $i$ and propagate uncertainties correctly, we implement a Markov Chain Monte Carlo (MCMC) model in \textsc{PySTAN}\footnote{\url{http://mc-stan.org/interfaces/pystan}}. We choose an isotropic prior for the inclination such that $\cos{i} \sim \text{Uniform}(0,1)$, $i \in [0^{\circ}, 90^{\circ}]$. The rotational frequency $f$ and radius $R$ are taken to be normally distributed. Finally, we model our projected velocities as $v\sin{i} \sim \text{Normal}(2 \pi R f \cdot \sin{i}, \sigma_{v\sin{i}})$, and we assume a 10 per cent uncertainty on $v\sin{i}$ in the value given by \citet{abt02}. We burn-in this model for 5000 MCMC steps, discard these, and generate a further 5000 samples from the posterior distribution. We find an inclination angle of $48\pm6^\circ$ and display a normalized histogram of these samples in Fig.~\ref{fig:rotations} as a marginal posterior probability density function, together with those of the other stars we shall discuss in turn. The inclination found is consistent with the orientations implied by the interferometric measurements of the circumstellar disc \citep{quirrenbach97,tycner05}.

The highest-amplitude frequency can therefore be explained as either an unresolved group of frequencies arising from SPB pulsation, or as the rotation frequency. We favour the pulsation explanation due to its ability to account for all the variability in the time series at all frequencies. Nevertheless, it is possible that the fundamental pulsation time scale is commensurate with the rotational time scale.

\subsection{Atlas}
Atlas (27\,Tauri, HR\,1178, HD\,23850) is the second-brightest member of the Pleiades ($V$\,=\,3.63\,mag). While there are a number of nearby stars separated by more than 50 arcsec \citep[identified as Atlas\,B--H in the WDS;][]{mason01}, the first report of Atlas\,A being a double star consisting of a 5\,mag primary (Atlas\,Aa) separated by 0.8\,arcsec from a 8\,mag secondary (Atlas\,Ab) was made by \citet{struve1837}. However, as reported by \citet{lewis1906}, apart from the first two measurements by F.~G.~W.~Struve in 1827 and 1830, multiple attempts by Struve and others failed to reobserve Atlas\,A as a double star until Lewis and W.~B.~Bowyer were able to resolve a secondary on consecutive nights in 1904 February. The WDS records a last `satisfactory' observation date for this component in 1929 and notes that the magnitudes, position angles and separations vary considerably, calling the existence of this companion into question \citep{mason01}. Nevertheless, subsequent references to its existence may still be found in the recent literature \citep[e.g.][]{neiner15}.

\citet{abt65} found Atlas to be a spectroscopic binary with a claimed orbital period of 1254\,d. However, with further measurements, this detection was also disputed \citep{pearce71}. 

Unambiguous confirmation that Atlas is a binary came with lunar occultation measurements \citep{mcGraw74,bartholdi75,deVegt76}. These measurements showed a much closer companion than suggested by the earlier, disputed observations. Further long-baseline optical interferometric and radial velocity measurements \citep{pan04,zwahlen04} have firmly established the orbital parameters and masses of the components, as well as provided an independent geometric distance to the Pleiades (132$\pm$4\,pc).

The primary component, identified as Atlas~Aa1 in the WDS \citep{mason01}, is a B8III star, and has been used as the standard star for its spectral type \citep{mkk43,jm53}. It has a mass of 4.74$\pm$0.25\,M$_\odot$ \citep{zwahlen04}. The secondary component, Atlas~Aa2, orbits the primary in an eccentric ($e$=0.2384$\pm$0.0063), 290.984$\pm$0.079\,d orbit with an inclination of $i=107.87\pm0.49^\circ$. It is a B8V star, with a mass of 3.42$\pm$0.25\,M$_\odot$ \citep{zwahlen04}.

Atlas\,A is identified as being He-weak in the catalogue of chemically peculiar stars by \citet{renson09}, although this identification is noted as being doubtful and it is not clear which component is implicated. We were unable to find an original reference to Atlas' chemical peculiarity in the literature. Chemical peculiarities are, however, commonly associated with structured magnetic fields and \citet{neiner15} have detected a magnetic field in Atlas\,Aa2 using spectropolarimetry, with a polar field strength of the order of 2\,kG.

Chemical peculiarity is also commonly associated with photometric variability arising from the rotation of surface spots. \citet{wraight12} searched for variability in stars identified as being chemically peculiar, including Atlas, using observations from the STEREO satellites. They flagged a potential frequency of 0.406\,d$^{-1}$, but blending and the low significance of the signal led them to doubt the reliability of this measurement. Photometric variability had been previously detected in Atlas by \citet{mcNamara85,mcNamara87}. He found a frequency of 0.546\,d$^{-1}$ between 1983--4, and 0.267\,d$^{-1}$ between 1985--6, although, as is common in ground-based observations, these measurements are affected by daily aliasing.

The K2 light curve and amplitude spectrum of Atlas are shown in the second row of Fig.~\ref{fig:res}; we clearly detect variability. The presence of the 13th mag star 2MASS J03491192+2403507, around 15~pixels from Atlas, dilutes the signal of Atlas in the halo light curve. We therefore use a 12~pixel radius mask to mitigate this effect.

Three prominent peaks are seen in the amplitude spectrum. The peak at 0.8236\,d$^{-1}$ is the second harmonic of the lowest peak at 0.4121\,d$^{-1}$. This lowest frequency coincides with the suspected period observed with STEREO. A third prominent peak occurs at 1.341\,d$^{-1}$. Several other low-amplitude peaks are also present in the amplitude spectra; a list of frequencies is given in Table\,\ref{tab:Atlas}. Most of these frequencies are harmonics of the two principal frequencies, but there is also a third independent frequency at 0.7627\,d$^{-1}$ and its harmonic, and a combination frequency of the two high-amplitude principal frequencies at 0.5092\,d$^{-1}$. The presence of this combination frequency reveals that this variability is caused by pulsations; such frequencies cannot be caused by rotation \citep[e.g.][]{kurtz15}.

It is not immediately clear whether these pulsations are present in Atlas\,Aa1 or Aa2, although we note that the amplitudes are diluted by the presence of the other star. Atlas\,Aa1 is approximately five times brighter than Atlas\,Aa2, so the pulsations will be diluted by a factor of 1.2 if they are present in the primary, and by a factor of 6 if they are present in the secondary. The large amplitude of two of the principal frequencies would seem to favour that they occur in the primary component, but further investigation including modelling the stars may provide greater clarity.

\subsection{Electra}

Electra (17\,Tauri, HR\,1142, HD\,23302) is the third-brightest star in the Pleiades ($V$\,=\,3.70\,mag), and is a blue giant of spectral type B6IIIe. While some lunar occultation measurements of Electra have been found to be consistent with a single star \citep{mcGraw74,eitter74}, others suggest it has a companion \citep{bartholdi75,deVegt76,richichi96}. \citet{abt65} found evidence that Electra was a spectroscopic binary with a period of $\sim$100\,d. However, while finding Electra may have a variable radial velocity, \citet{pearce71} were unable to confirm the period claimed by \citet{abt65}. \citet{jarad89} also claim Electra may be a spectroscopic binary, although with a period of only $\sim$4\,d.

Previous photometric observations of Electra have found it to be constant within the precision of the measurements. Subsequently, \citet{mcNamara85,mcNamara87} used Electra as the reference star for differential photometry with other members of the Pleiades.

The time series and amplitude spectrum of Electra are shown in the third row of Fig.~\ref{fig:res}. Electra has the lowest amplitude variability of any of the stars considered in this paper. Low-amplitude frequency groupings are present in the ranges 0.8--1\,d$^{-1}$ and 1.8--2\,d$^{-1}$. These are SPB pulsations showing the same type of structure that was previously discussed for Alcyone, with several base frequencies and combination frequencies \citep{kurtz15}. The detected frequencies, their amplitudes, and phases are provided in Table~\ref{tab:Electra}. We are able to identify several possible combination frequencies. However, as was the case for Alcyone, further analysis of Electra is limited by the resolution of the K2 time series and the closely-spaced frequencies within each group.

\subsection{Maia}\label{sect:maia}

\begin{figure*}
	\includegraphics[width=2.1\columnwidth]{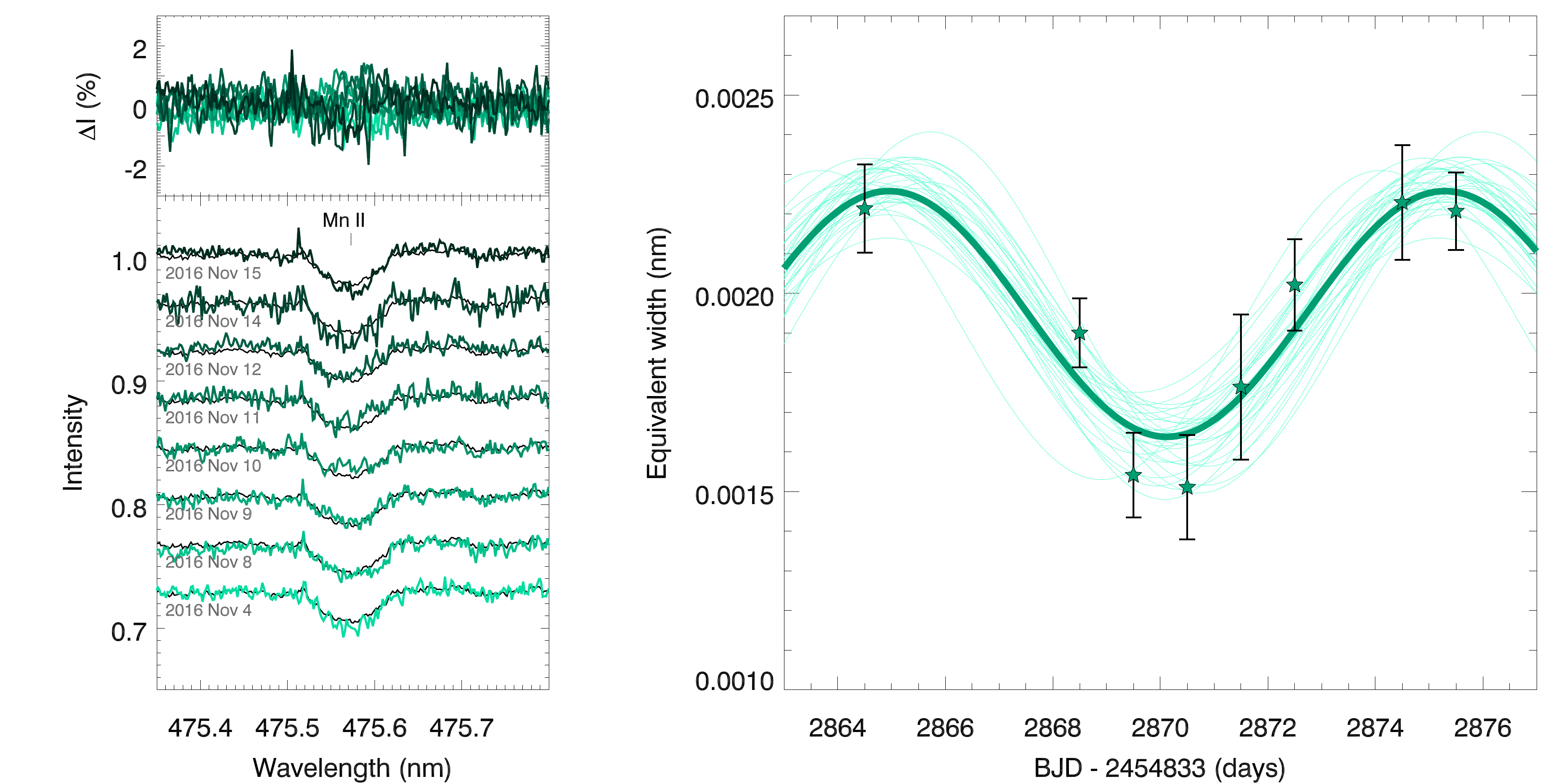}
	\caption{Variability in the 475.6\,nm \ion{Mn}{ii} line of Maia in spectra taken with the Hertzsprung SONG telescope. \emph{Left} Section of the spectrum around the line. The lower panel shows the spectrum for each night, offset, with time increasing from bottom (light green) to top (dark green). The black lines show the average spectrum. The top panel shows the difference from the average spectrum. \emph{Right} Equivalent width of the line as a function of time. The thick solid line shows the best fit of a single sinusoid to the observations (green stars), while the faint background lines show draws from the MCMC posterior distribution. The best-fitting sinusoid from the MCMC fit to the equivalent widths has a period of $10.34 \pm 0.08$\,d, consistent with the 10.29\,d found from the K2 photometry.}
	\label{fig:maia_spect_Mn}
\end{figure*}

Maia (20\,Tauri, HR\,1149, HD\,23408), a B7III star, is the fourth-brightest member of the Pleiades ($V$\,=\,3.87\,mag). The uniqueness of its relatively narrow-lined spectrum amongst the B stars in the Pleiades spurred early interest in detecting variability, although a clear detection proved elusive.

\citet{adams04} was first to claim variability in radial velocity measurements, suspecting binarity. \citet{henroteau1920} found tentative evidence of variability with a period of about 2\,h in the \ion{Mg}{ii} line at 448.1\,nm, and raised the possibility of pulsations. However, other early measurements failed to find variability in radial velocity \citep{merrill1915} or photometry \citep{guthnick1922}. \citet{walker52} found a suggestion of photometric variability, again with a period of about 2\,h. \citet{struve1955} went further, not only suggesting that Maia exhibited radial velocity variability on a 1--4\,h time-scale, but hypothesised a class of similar variable stars, lying in the HR diagram between the $\beta$~Cephei and $\delta$~Scuti pulsators, which he called the Maia variables. A more detailed radial velocity and photometric study of Maia by \citet{struve1957}, however, did not support the existence of short-period variations in radial velocity or brightness. Instead, it was found that the helium lines varied in intensity and width in a non-periodic manner, at intervals of the order of 2\,h. Nevertheless, the idea of the Maia variable class remained.

Further searches for variability in Maia have been negative \citep{breger72,percy78,mcNamara85,mcNamara87,percy2000,paunzen2013}. Other candidate Maia variables have been investigated but, although several stars with variability on the appropriate time-scale were found, clear evidence of the existence of Maia variables as a single, distinct class had not been forthcoming \citep[e.g.][]{lehmann95,scholz98,weiss98,kallinger02,kallinger04,deCat07}.

\citet{aerts05} suggested that the Maia variable class was unnecessary; any variable that matched the description could be a rapidly rotating SPB star with the gravity modes shifted to higher frequency by the Coriolis force \citep{townsend03,townsend05b}. \citet{savonije05} and \citet{townsend05} have also investigated Rossby waves, which are able to be excited in models of rotating SPB stars.

Recent results have firmly established that there are short-period pulsators, both within the SPB instability strip and between the SPB and $\delta$~Scuti instability strips \citep{degroote09,mowlavi13,mowlavi16,lata14,balona15,balona16}. \citet{salmon14} have argued that models of rapidly rotating SPB pulsators are consistent with these observations. \citet{mowlavi16} do find that this class of pulsators in the cluster NGC\,3766 are rapidly rotating, but have dubbed them fast-rotating pulsating B (FaRPB) stars while their behaviour is under further investigation. Others are sticking to the historic ``Maia variable'' name for this class \citep[e.g.][]{balona15,balona16}.

Despite the ongoing debate over the existence of Maia variables, we can at least now settle the question of Maia's variability. The K2 time series and amplitude spectrum of Maia are shown in the fourth row of Fig.~\ref{fig:res}. For the first time, we have unambiguously detected variability in Maia. One clear frequency is found in the amplitude spectrum at 0.0967$\pm$0.0008\,d$^{-1}$; the variability has a 10\,d period. Its Fourier parameters are given in Table~\ref{tab:Maia}. Other low-amplitude peaks are present at similar periods, including at the harmonic of the primary frequency, but because these time-scales are similar to the instrumental long-term trends that were removed, we are cautious to not treat these low-amplitude modes as necessarily being intrinsic to the star. Nevertheless, this much is clear: Maia is variable, but Maia is \emph{not} a Maia variable. 

Maia is known to be chemically peculiar. \citet{struve1933} reported that he found the spectrum particularly difficult to characterize. He eventually came to the conclusion, based off the small projected rotational velocity and the apparently inconsistent temperatures between relatively weak helium lines and relatively strong metallic lines, that Maia was a rapidly rotating star seen pole-on \citep{struve1945}. Amongst the metal lines, the ultraviolet \ion{Mn}{ii} lines were especially strong. While \citet{struve1945} left open the possibility of slow rotation with a peculiar spectrum, the pole-on hypothesis was supported by \citet{huang56}. When the realization that there were whole classes of chemically peculiar stars came, Maia was soon identified as belonging to the He-weak (or CP4) stars \citep{searle64,jaschek69}. Subsequently Maia was considered amongst the HgMn (CP3) stars; \citet{wolff78} noted intermediate-strength Mn lines relative to HgMn stars, but no Hg lines, and did not include Maia on their list of HgMn stars. Nevertheless, \citet{heacox79} included Maia in their sample of HgMn stars for which they measured detailed abundances; again Hg was not detected. It would appear that Maia presents an intermediate case between the two classes. \citet{renson09} list Maia as a He-weak Mn star in their catalogue.

Many HgMn stars show evidence of line-profile variations caused by the rotation of chemical surface spots \citep[e.g.][]{ryabchikova99,adelman02}. Rotational modulation has also been observed in the light curves of other HgMn stars \citep[e.g.][]{balona11}. We suggest that the photometric variability of Maia is similarly caused by rotational modulation resulting from a large chemical spot.

\subsubsection{SONG spectroscopy of Maia} \label{sect:maiaspect}
To confirm the existence of a chemical spot on Maia we obtained high-resolution spectra with the automated Hertzsprung SONG 1\=/m~telescope at Observatorio del Teide on eight nights between 2016~November~4 and 15 to cover an entire rotation period. SONG (Stellar Observations Network Group) is a planned global network of 1-m telescopes, of which the Hertzsprung SONG telescope is the first node. The telescope is equipped with a coud\'e \'echelle spectrograph with an iodine cell and a ThAr lamp (Grundahl et al., in prep.). The first SONG results, asteroseismology of the G5 subgiant $\mu$~Herculis, were recently presented \citep{grundahl17}.

Each observation of Maia consisted of a 180\=/s exposure, of which three were taken in succession on each night. There are 51 orders in the \'echelle spectra, covering a range of 440--690\,nm with a spectral resolution of $R=90\,000$. Bias and flat-field calibration frames were taken before the beginning of each observing night, while ThAr spectra were measured before and after the observations for wavelength calibration. The spectra were reduced using a C++ based pipeline using the implementation of the routines of \citet{piskunov02} by \citet{ritter14}. For each exposure, the barycentric Julian mid-time and barycentric velocity correction were made using the program \textsc{BarCor}\footnote{\url{sirrah.troja.mff.cuni.cz/~mary}} by M. Hrudkov\'a.

Spectral lines were identified by reference to the VALD database \citep{kupka99,ryabchikova15}, assuming stellar parameters $T_\mathrm{eff}=12550$\,K
(this work, see Section~\ref{sect:maiainterf}), $\log g=3.5$, and the detailed chemical abundances of \citet{heacox79}. A single spectrum of each night was obtained from the sum of the three exposures. Given the rotation period of $\sim$10\,d, combining consecutive spectra should not smear out the signal. The time series of the spectra were searched for line profile variations. Several lines show clear evidence of line profile variability, however the most prominent cases involve blended lines. Fig.~\ref{fig:maia_spect_Mn} shows the relatively weak variability in an isolated \ion{Mn}{ii} line. To quantify the variability in this line, we used the equivalent width as a proxy for the line profile shape, fitting the line with a Gaussian function. Uncertainties in the measurements were determined using Monte Carlo simulations with the scatter in the surrounding continuum spectrum used to draw realizations of the spectrum. The time series of the equivalent width measurements is shown in the right panel of Fig.~\ref{fig:maia_spect_Mn}, which shows clear variability over the observations consistent with a 10\,d period. We determined the period of this variability with a Markov Chain Monte Carlo fit of a sinusoid to the time series. With only a handful of data points sampling a single period of the variation, we constrain our model with a normal prior with mean and uncertainty taken from the K2 time series, allowing phase, amplitude and mean to vary with uniform priors. We find an excellent agreement between the period and phase of the photometric and spectroscopic variability, supporting the conclusion that the variability of Maia is caused by a chemical spot. A more detailed analysis of the spectra would allow for Doppler mapping of the chemical spot, however this is beyond the scope of this paper.

\subsubsection{CHARA interferometry of Maia}\label{sect:maiainterf}

We can determine the inclination angle of Maia by combining the rotation frequency (0.0962$\pm$0.0002\,d$^{-1}$) with the projected rotational velocity \citep[$v \sin i = 33\pm5 \mathrm{km\,s}^{-1}$;][]{royer02} and radius. 

To determine the radius of Maia, we have conducted interferometric observations with the PAVO beam combiner \citep{ireland08} at the CHARA Array at Mt. Wilson Observatory, California \citep{tenbrummelaar05}. The CHARA Array consists of six 1-m telescopes arranged in a Y-shaped configuration, with baselines ranging from 30--330\,m. PAVO is a visible wavelength ($\sim$600--900\,nm) pupil-plane beam combiner. Although PAVO can combine three beams, we have used the two-telescope mode, for which calibration is more accurate.

Observations were made with three different baselines on three nights in November~2013~and~2015. A summary of the observations is provided in Table~\ref{tab:pavoobs}. Calibration of the target fringe visibilities is made through comparison to the fringe visibilities of calibrator stars. Ideally, these stars are as small as possible to be unresolved by the interferometer, while being bright and close ($<10^\circ$) to the target. The calibrator stars are listed in Table~\ref{tab:pavocals}. Their expected angular diameters were calculated as the average of values obtained from the ($V-K$) relations of \citet{kervella04} and \citet{boyajian14}. The stars were observed in the sequence \emph{calibrator 1} -- \emph{Maia} -- \emph{calibrator 2}, with two minutes of visibility data collected for each object.

\begin{figure}
	\includegraphics[width=1\columnwidth]{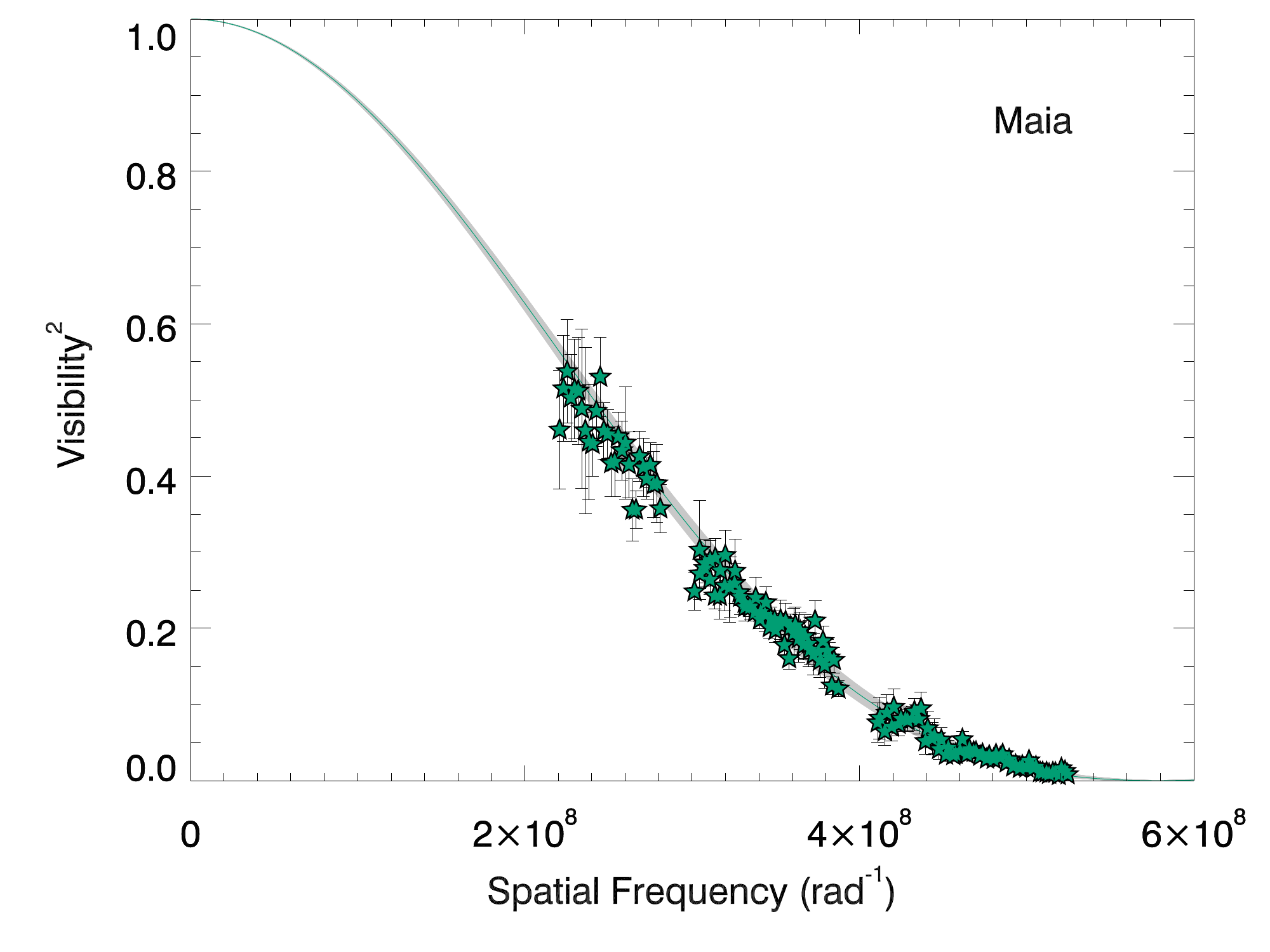}
	\caption{Interferometric measurements of Maia. Green stars show fringe visibility measurements made with the PAVO beam combiner at the CHARA array. The curve shows the best-fitting limb-darkened disc model.}
	\label{fig:maia_interf}
\end{figure}

The data were reduced and analysed using the PAVO reduction pipeline, which has been well-tested and used for previous studies \citep[e.g.][]{bazot11,derekas11,huber12,maestro13,white13}. The calibrated squared visibility measurements are shown in Fig.~\ref{fig:maia_interf}.

\begin{table}
 \centering
 \begin{minipage}{140mm}
 \caption{Log of PAVO interferometric observations of Maia}
 \label{tab:pavoobs}
 \begin{tabular}{@{}lccc@{}}
     \hline%\hline
 \textsc{ut} Date & Baseline\footnote{The baselines used have the following lengths: \\S2W2, 177.45\,m; S2E2, 248.13\,m; E1S1, 330.66\,m.} & No. of scans & Calibrators\footnote{Refer to Table~\ref{tab:pavocals} for details of the calibrators.}\\
 \hline
 2013 November 9 & E1S1 & 2 & b   \\
 2015 November 8 & S2W2 & 2 & ad  \\
 2015 November 9 & S2E2 & 3 & ace \\
 \hline
 \end{tabular}
 \end{minipage}
 \end{table}

 \begin{table}
 \centering
 \caption{Calibration stars used for interferometric observations}
 \label{tab:pavocals}
 \begin{tabular}{lccrccc}
 \hline%\hline
 HD & Sp. type & $V$ & $V-K$ & $E(B-V)$ & $\theta_{V-K}$ & ID \\
 \hline
 21050  & A1V   & 6.070 & $-$0.072 & 0.036 & 0.198 & a \\
 23923  & B8V   & 6.172 & $-$0.044 & 0.056 & 0.191 & b \\
 23950  & B8III & 6.070 &    0.094 & 0.051 & 0.218 & c \\
 25175  & A0V   & 6.311 &    0.237 & 0.120 & 0.209 & d \\
 27309  & A0p   & 5.345 & $-$0.291 & 0.000 & 0.244 & e \\
 \hline
 \end{tabular}
 \end{table}

The calibrated fringe visibilities were fit with a linearly limb-darkened disc model, given by \citep{hanburybrown74}
\begin{equation}
V = \left( \frac{1-u}{2} + \frac{u}{3} \right)^{-1}\left[ (1-u) \frac{J_1(x)}{x} + u (\pi/2)^{1/2} \frac{J_{3/2}(x)}{x^{3/2}} \right], \label{eqn:ldvis}
\end{equation}
where
\begin{equation}
x \equiv \pi B \theta_\mathrm{LD} \lambda^{-1},\label{eqn:ldvis_x}
\end{equation}
and $V$ is the visibility, $u$ is the wavelength-dependent linear limb-darkening coefficient, $J_n(x)$ is the $n^\mathrm{th}$
order Bessel function of the first kind, $B$ is the projected baseline, $\theta_\mathrm{LD}$ is the angular diameter after correction for
limb-darkening, and $\lambda$ is the wavelength at which the observations was made. The quantity $B\lambda^{-1}$ is
referred to as the spatial frequency.

To determine the value of the limb-darkening coefficient, we use the grid of coefficients calculated by \citet{claret11} for \textsc{atlas} atmosphere models. We interpolate the grid to the spectroscopic values of $\log g = 3.5$ and $T_\mathrm{eff} = 13300\,K$, as found by \citet{heacox79}. Our linear limb-darkening coefficient is $u=0.31\pm0.04$. We note that the \textsc{atlas} models in the grid have scaled solar compositions, whereas Maia is chemically peculiar, however the dependence of metallicity on the model limb-darkening coefficient is minimal in this region of the parameter space. There is further uncertainty in the impact of the suspected large chemical spot on the effective limb darkening. We adopt a relatively large uncertainty in the limb-darkening coefficient in order to account for these systematic uncertainties. Because Maia is a relatively slow rotator, we do not take into account possible oblateness and gravity darkening.

We determine the uncertainty in the angular diameter using Monte Carlo simulations that take into account the uncertainties in the visibility measurements, the wavelength calibration (5\,nm), the calibrator sizes (5 per cent), and the limb-darkening coefficient. The uniform disc diameter (i.e. $u=0$) is $\theta_\mathrm{UD}=0.438\pm0.005$\,mas. With our adopted linear limb-darkening coefficient, we find $\theta_\mathrm{LD}=0.451\pm0.006$\,mas. The best fitting limb-darkened disc model is shown as the curve in Fig.~\ref{fig:maia_interf}.

Adopting the distance to the Pleiades as determined from VLBI parallax measurements \citep[$136.2\pm1.2$\,pc;][]{melis14}, we find a radius for Maia of $6.61\pm0.11\mathrm{R}_\odot$. Combining our angular diameter with the estimate of bolometric flux by \citet{vanBelle08}, for which we adopt a 5 per cent uncertainty, we find an effective temperature of $T_\mathrm{eff} = 12550\pm150$\,K.

The calculation of the inclination angle from the radius, projected rotational velocity, and rotation period reveals that Maia is seen close to equator-on. We implemented the same MCMC model that we previously used to determine the possible inclination of Alcyone in Section~\ref{sect:resalcyone} to properly account for the prior distribution of $i$ and propagate uncertainties correctly. The resulting posterior distribution is shown in Fig.~\ref{fig:rotations}. We find a range of possible inclinations, with a 99 per cent probability of it being above 42$^\circ$.

\subsection{Merope}
Merope (23\,Tauri, HR\,1156, HD\,23480) is the fifth-brightest member of the Pleiades ($V$\,=\,4.18\,mag) and is of spectral type B6IVe. 
Merope appears single in lunar occultation \citep{mcGraw74,eitter74,deVegt76} and speckle interferometry measurements \citep{mason93}.

Previous photometric observations of Merope have revealed high-amplitude variability. \citet{mcNamara85,mcNamara87} found a consistent frequency of 2.04\,d$^{-1}$ across several years. This frequency is, coincidentally, half the thruster-firing frequency for the K2 mission. The K2 time series and amplitude spectrum of Merope are shown in the fifth row of Fig.~\ref{fig:res}. The previously-known frequency at 2.04\,d$^{-1}$ is clearly recovered. The halo photometry method works remarkably well in this situation, preserving the stellar variability despite the unfortunate similarity to the thruster firing period, whereas other methods that attempt to detrend the telescope pointing drift signal would likely overfit and remove part of the stellar signal as well. While there are small peaks at 4.08 and 6.13\,d$^{-1}$ in the amplitude spectrum, it is unclear if these are harmonics of the primary frequency, or a residual of the pointing drift.

The width of the peak in the amplitude spectrum is indicative of there being several closely spaced periods or a non-coherent signal, resulting in the beating apparent in the time series. There are several possible explanations for the variability. Merope is a Be star so it has a circumstellar disc that may exhibit variability. Alternatively, the variability may have a stellar origin, either as rotation, with differential rotation accounting for incoherent variability, or from SPB pulsations, with several closely spaced $g$-mode frequencies.

Rotation may be ruled out as the cause of variability by considering the rotational velocity this would require. Using the radius we previously determined (4.79$\pm$0.17\,R$_\odot$) with the mass determined by \citet{zorec12} of 4.25$\pm$0.08\,M$_\odot$, we find that the critical rotation frequency of Merope is 1.39$\pm$0.07\,d$^{-1}$. The observed variability occurs well above this frequency, and therefore we can be confident that rotation is not the cause.

The frequency of the variability in Merope has been constant over the course of 30\,yr. Expecting that variability in the circumstellar disc would be less stable, we believe pulsations provide the most plausible explanation. We note, however that the oscillation spectra of these presumed SPB pulsations are qualitatively quite different from those seen in Alcyone and Electra, as well as Taygeta, which we discuss next, both in terms of amplitude and the spacing between modes. The oscillations in Merope show more similarity with pulsations detected in other Be stars from the ground, including the beating between several frequencies, which has been advanced as a possible explanation for the production of the decretion disc \citep{rivinius03,kee14,baade16}. Of course, the SPB pulsations seen in Alcyone, Electra, and Taygeta are at such a low amplitude that they would not be observed from the ground.

To resolve the individual frequencies, a significantly longer time series is required. Although a list of frequencies is provided in Table~\ref{tab:Merope}, there are frequencies more closely spaced than the formal resolution of the time series, and consequently the Fourier components are unreliable \citep{loumos78}. Unfortunately, K2 will not return to the Pleiades during the remaining campaigns, however, with its large amplitude, Merope is a suitable candidate for observations by the BRITE-Constellation of nanosatellites \citep{weiss14}.

\subsection{Taygeta} 
Taygeta (19\,Tauri, HR\,1145, HD\,23338) is the sixth-brightest star in the Pleiades ($V$\,=\,4.30\,mag), and is of spectral type B6IV. \citet{abt65} reported it to be a spectroscopic binary with a period of 1313\,d, however \citet{pearce71} were unable to confirm this companion with more data. \citet{liu91} found the radial velocity to be variable. Several lunar occultation \citep{bartholdi75b,deVegt76,fan91} and speckle interferometry measurements  \citep{mcAlister78,hartkopf84,mason93} had found either weak or no evidence of a companion. A faint companion was finally unambiguously detected in lunar occultation measurements by \citet{richichi94}.

\citet{mcNamara85} found no evidence of photometric variability in Taygeta, and \citet{adelman01} found it to be amongst the least variable stars observed by Hipparcos.

The K2 time series and amplitude spectrum of Taygeta are shown in the sixth row of Fig.~\ref{fig:res}. The light curve shows low-amplitude variability, consisting of several frequencies. The amplitude spectrum bears a striking resemblance to that of Electra, albeit scaled to a lower frequency, with a frequency grouping in the range 0.4--0.7\,d$^{-1}$, and another group in the range 1.2--1.4\,d$^{-1}$. We attribute this variability to SPB pulsations, with combination frequencies. A low-amplitude peak is also present at 19.9\,d. A list of the detected frequencies is provided in Table~\ref{tab:Taygeta}.

\subsection{Pleione} 
Pleione (28\,Tauri, HR\,1180, HD\,23862), the seventh-brightest member of the Pleiades ($V$\,=\,5.19\,mag), has spectral type B8Vne. Since first being identified as a Be star \citep[A.~C.~Maury, reported by][]{pickering1889}, Pleione has attracted a lot of attention due to its regular transitions between Be and shell star phases \citep[e.g.][]{frost1906,gulliver77,goraya90}. Pleione cycles between its Be and shell phases with a period of 34.5\,yr. The most recent transitions have been from shell to Be in 1988 \citep{sharov88}, and Be to shell in 2006 \citep{tanaka07}. \citet{hummel98} suggested that such transitions are due to the precession of the disc, likely under the influence of a misaligned binary orbit. Polarimetric observations of Pleione by \citet{hirata07} found the intrinsic polarization angle varied from 60$^\circ$ to 130$^\circ$ between 1974 and 2003, providing evidence of changes in the disc axis.

Pleione has been monitored for radial velocity variations for many years \citep{struve43,merrill52,ballereau88,katahira96}, which has revealed it to be a multiple system. The nearest companion has a low mass and has an orbital period of 218\,d \citep{nemravova10}, and it has been suggested this is the source of the precession of the disc \citep{hirata07}. The next furthest out has a period of $\sim$35 years, which is commensurate with the shell--Be transition period, leading to it also being associated with this process \citep{harmanec82,luthardt94}. This companion is identified as being CHARA 125, which was detected using speckle interferometry \citep{mcalister89,mason93} with a separation of 0.23\,arcsec. That it is not always detected in speckle measurements suggests a large magnitude difference of $\Delta V \sim 3.5$\,mag. \citet{gies90} measured the extent of Pleione's circumstellar envelope by observing the H\,$\alpha$ emission line during a lunar occultation and found the envelope to be asymmetric, which they proposed could be due to the presence of this speckle binary companion. 

The transition between Be and shell phases is accompanied by photometric variability \citep{calder37,binnendijk49,sharov76,tanaka07}. Shorter time-scale variability was detected by \citet{mcNamara87}, who identified a primary frequency at either 1.21~or~2.21\,d$^{-1}$, with a second frequency at 1.18\,d$^{-1}$.

The K2 time series and amplitude spectrum are shown in the lowest panel of Fig.~\ref{fig:res}. The variability bears a strong resemblance to that of Merope, with a large amplitude and closely spaced frequencies at $\sim$1.7\,d$^{-1}$. Other frequencies present at 0.41, 0.82 and 1.34\,d$^{-1}$ are those of nearby Atlas, which contaminates the light curve. A low-amplitude peak is also present at 7.46\,d$^{-1}$. There is no evidence of power at the frequencies reported by \citet{mcNamara87}. It may be the case that the variability has changed in the intervening years, during which Pleione has gone through a Be period. Variations in frequency and amplitude have been previously seen in Be stars \citep{huat09}. However, the frequencies present in the K2 data are precisely 0.5\,d$^{-1}$ between the possible primary frequencies reported by \citet{mcNamara87}; with a particularly diabolical spectral window, the frequencies found by \citet{mcNamara87} may be aliases of the true frequencies at 1.7\,d$^{-1}$. A peak at $\sim$1.7\,d$^{-1}$ can be seen in the power spectrum of Pleione in figure 1 of \citet{mcNamara87}.

As for Merope, there are several possible interpretations for the observed variability. The variability may arise from the circumstellar disc, from rotational modulation, or from SPB pulsations. 

We consider it most likely that the cause of the variability is the same as what we have determined for Merope, namely SPB pulsations with several closely-spaced modes. Beating between these modes may provide the additional angular momentum to eject material into the circumstellar decretion disc, as was proposed as the origin of the Be phenomenon by \citet{rivinius98}. As is the case for Merope, the high amplitude and small spacing between modes are in contrast to what is observed in the oscillation spectra of Alcyone, Electra, and Taygeta. 

We are, however, unable to entirely rule out rotation as the cause of Pleione's variability through consideration of the critical velocity. Using the estimated mass and radius listed in Table~\ref{tab:prop}, we find a critical rotational frequency of 1.62$\pm$0.10\,d$^{-1}$, which is commensurate with detected frequencies. If the variability is caused by rotation, then Pleione is rotating very close to its critical velocity. We have calculated the inclination angle this would imply, as we have done for Alcyone and Maia, finding $i=53\pm5^\circ$; the posterior probability distribution is shown in Fig.~\ref{fig:rotations}.

While a list of indicative frequencies is provided in Table~\ref{tab:Pleione}, a longer time series is required to properly resolve individual frequencies and provide a fuller understanding of Pleione's variability. Of particular interest would be to investigate how the variability may change over the course of the shell--Be cycle. Fortunately, this variability has a large amplitude, making Pleione a suitable candidate for observations by the BRITE-Constellation of nanosatellites \citep{weiss14}.

\section{Source Code}
We are committed to open science, and have made the software presented in this paper available open-source. The main halo algorithm, \textsc{halophot}, is available at \url{https://github.com/hvidy/halophot}. We invite interested users to apply this to other data and contribute to its ongoing development. The \textsc{Jupyter} notebook used for estimating rotational inclinations with \textsc{PyStan} MCMC is available at \url{https://github.com/benjaminpope/inclinations}. Finally, the \textsc{k2sc} algorithm used for finer corrections \citep{k2sc}, is also available at \url{https://github.com/OxES/k2sc}. All code is provided under a GPL v3 license. 

All original light curves discussed in this paper will be made available on the Mikulski Archive for Space Telescopes (MAST) as High-Level Science Products, and through the \textit{Kepler} Asteroseismic Science Operations Centre (KASOC) database \footnote{\url{http://kaosc.phys.au.dk}}.

\section{Discussion and Conclusions}

We have presented a new method for simultaneous photometry and systematics correction of saturated stars with K2, and used this method to obtain the first high-precision space-based light curves of the seven brightest stars in the Pleiades. Halo photometry dramatically reduces the number of pixels required to observe bright stars; using a 20~pixel radius masks, this is equivalent to approximately six to twelve $Kp$\,=12\,mag stars. Good precision can be achieved using masks as small as 12~pixels in radius. Halo photometry therefore extends the dynamic range of K2 to include naked-eye stars, allowing for the combined use of photometric time series with other astronomical tools, including high-resolution spectroscopy, interferometry and polarimetry, to form a more complete understanding of these stars and stellar systems. 

The light curves of the bright Pleiades stars show a wide range of variability. Low amplitude SPB pulsations and combination frequencies have been detected in Alcyone, Electra and Taygeta. Pleione and Merope have high amplitude oscillations that may also be due to SPB pulsations, however longer time series are necessary to resolve individual modes; this may be achieved with the BRITE-Constellation of nanosatellites \citep{weiss14}. In contrast, the high amplitude oscillations of Atlas, which we believe are most likely to originate from the primary component, are well-resolved. 

The combination of pulsations and rotation at near-critical velocity has been invoked to explain the production of circumstellar decretion discs in Be stars \citep{rivinius98,rivinius03,huat09,kee14,baade16}. It is interesting to note, then, any differences in the pulsations seen in Be and normal B stars. The beating between closely-spaced frequencies and high-amplitude variability seen in the Be stars Merope and Pleione would seem to be consistent with this hypothesis. The normal B star Atlas\,Aa1, although having high-amplitude pulsations and a projected rotational velocity of 240\,km\,s$^{-1}$, does not show any beating between frequencies. The more-evolved Be stars Alcyone and Electra have low-amplitude oscillation spectra that have more in common with the normal B star, Taygeta. While Alcyone and Electra were not observed to have large amplitudes now, other Be stars, such as the CoRoT target HD\,49330 have exhibited g-modes that grow in conjunction with an outburst \citep{huat09}. The reason why Taygeta is not a Be star may be due to rotation; of the three, Taygeta has the lowest projected rotational velocity. Whether the differences in $v\sin i$ are due to intrinsic differences in rotation rates or the effect of projection may be resolved by our ongoing campaign to measure the angular diameters and the oblateness of these stars with the CHARA Array. 

For Maia we have combined the photometric time series with high-resolution spectroscopy and interferometry to determine that Maia is viewed equator-on, with variability due to large chemical spots in the photosphere. With a 10\,d period, we have conclusively determined that Maia is not a so-called Maia variable. Now that the variability of Maia has been detected and it is not on hour time scales, further reference to Maia variables will become increasingly confusing. For further studies that are conducted into hour-scale variability in late B stars, we implore, for the sanity of future astronomers, that at the very least they no longer be referred to as Maia variables.

The halo photometry method is very general in its formulation and applicability, and will therefore be also of interest in planning for the Transiting Exoplanet Survey Satellite (TESS), which will cover nearly the whole sky in a succession of 27\,d pointings to search for exoplanets transiting bright stars. TESS will provide full-frame images at a 30\,min cadence, and postage stamps of individual targets at 2\,min or 20\,s cadence. As with \textit{Kepler} and K2, competition for this pixel allocation will place pressure on bright targets with long bleed columns. In addition to this, detector nonlinearity is expected to set in at $\sim 30$~times the single-well depth. With minimal saturation for $V\lesssim$\,6.2--6.7\,mag depending on how the PSF lands on the pixels, this would suggest that simple aperture photometry of stars much brighter than $V \sim$\,3--4\,mag will not be possible even with apertures that capture the entire bleed column (Jon Jenkins \& Roland Vanderspek, priv. comm.). It will therefore be necessary to use halo photometry for any TESS targets brighter than this magnitude, or targets unfavourably close to the edge of the detector, and to economize on pixel allocation. A valuable future step will be to simulate the TESS PSF, saturation, and scattered light, to determine the required halo masks both for asteroseismology and for transiting exoplanet searches. The ability to observe very bright stars with TESS raises the prospect of searching for planets transiting some of the nearest main-sequence stars to the Solar System, which being so bright, are the most suitable systems for transit spectroscopy.

In the particular case of B-type stars, there are two major lines of enquiry that may be pursued with TESS: delineating the borders of the instability strips and detailed seismic modelling. The predicted locations of the instability strips rely on models for which the input physics is still uncertain. Changes to the iron opacity, for example, have been shown to shift the location of the instability strips \citep[e.g.][]{moravveji16b}. Observationally defining the borders of the SPB instability strip requires the detection of at least two independent frequencies in many well-characterized stars. This does not require particularly long time series, and can be achieved for stars observed by TESS for only 27\,d, as well as those observed from the ground and with K2. Detailed asteroseismic modelling, however, requires well-resolved frequencies, which for SPB stars requires long time series. \textit{Kepler} observations of SPB stars have enabled constraints on mixing due to convective-core overshooting \citep{papics14} and diffusion \citep{moravveji15}, as well an inversion of the internal rotational profile of a star \citep{triana15}. Such work can be continued by TESS for stars that are within the continuous viewing zone, and will subsequently be observed for a full year. Halo photometry will allow for these observations to be made for the brightest B stars that will be observed by TESS.

It may be important to consider the halo method in the context of other upcoming ground-based and space-based photometry missions, such as PLATO \citep[PLAnetary Transits and Oscillations of stars;][]{rauer14}, JWST \citep[James Webb Space Telescope;][]{gardner06,2014PASP..126.1134B}, NGTS \citep[the Next Generation Transit Survey;][]{ngts}, or CHEOPS \citep[CHaracterising ExOPlanets Satellite;][]{cheops}. In each of these cases, it may be possible to enhance the photometry of a defocused PSF with or without saturation, for asteroseismology or exoplanetary science. In this regime, TV minimization, or a similar algorithm, may improve the performance of any such instrument spreading astrophysical signal over many different pixel realizations. We note that we do not yet have a consistent explanation of why TV, as opposed to relatives such as QV, is so much more effective, and that there may be new insights to be gained from theoretical studies, that may expose limitations on the method or reveal improvements. We encourage other groups to refine the halo method and consider the applicability of Total Variation more generally in time-series analysis.

\section*{Acknowledgements}
The authors would like to thank the entire K2 team, without whom these results would not be possible. We would like to thank Jon Jenkins and David Armstrong for their very helpful comments on early presentations of this work. 

This paper includes data collected by the K2 mission. Funding for the K2 mission is provided by the NASA Science Mission directorate. This work also includes observations made with the Hertzsprung SONG telescope operated at the Spanish Observatorio del Teide on the island of Tenerife by the Aarhus and Copenhagen Universities and by the Instituto de Astrof\'isica de Canarias. This work is also based upon observations obtained with the Georgia State University Center for High Angular Resolution Astronomy Array at Mount Wilson Observatory. The CHARA Array is supported by the National Science Foundation under Grants No. AST-1211929 and AST-1411654. Institutional support has been provided from the GSU College of Arts and Sciences and the GSU Office of the Vice President for Research and Economic Development.

Funding for the Stellar Astrophysics Centre is provided by The Danish National Research Foundation. The research was supported by the ASTERISK (ASTERoseismic Investigations with SONG and Kepler) and MAMSIE (Mixing and Angular Momentum tranSport in massIvE stars) funded by the European Research Council (Grant agreements no.: 267864 and 670519, respectively). TRW and VSA acknowledge the support of the Villum Foundation (research grant 10118). TRW, TRB and MBN acknowledge the support of the Group of Eight universities and the German Academic Exchange Service through the Go8 Australia-Germany Joint Research Co-operation Scheme. BP is grateful for the financial support of the Clarendon Fund and Balliol College. PIP acknowledges support from The Research Foundation -- Flanders (FWO), Belgium. DH acknowledges support by the Australian Research Council's Discovery Projects funding scheme (project number DE140101364) and support by the NASA Grant NNX14AB92G issued through the \textit{Kepler} Participating Scientist Program. PGB acknowledges the ANR (Agence Nationale de la Recherche, France) program IDEE (n$^\circ$ANR-12-BS05-0008) ``Interaction Des Etoiles et des Exoplanetes'' and also received funding from the CNES grants at CEA.

This research made use of NASA's Astrophysics Data System; the SIMBAD database, operated at CDS, Strasbourg, France; the VALD database, operated at Uppsala University, the Institute of Astronomy RAS in Moscow, and the University of Vienna; the Washington Double Star Catalog maintained at the U.S. Naval Observatory; and the BeSS database, operated at LESIA, Observatoire de Meudon, France. Data presented in this paper were obtained from the Mikulski Archive for Space Telescopes (MAST). STScI is operated by the Association of Universities for Research in Astronomy, Inc., under NASA contract NAS5-26555. Support for MAST for non-HST data is provided by the NASA Office of Space Science via grant NNX13AC07G and by other grants and contracts. 

This research made use of Astropy, a community-developed core Python package for Astronomy \citep{2013A&A...558A..33A}; \texttt{ds9}, a tool for data visualization supported by the Chandra X-ray Science Center (CXC) and the High Energy Astrophysics Science Archive Center (HEASARC) with support from the JWST Mission office at the Space Telescope Science Institute for 3D visualization; the IPython package \citep{PER-GRA:2007}; \texttt{matplotlib}, a Python library for publication quality graphics \citep{Hunter:2007}; as well as \textsc{SciPy} \citep{jones_scipy_2001}, and the Astronomy Acknowledgement Generator.

%The authors would like to acknowledge the Gadigal people of the Eora nation and the Norongerragal and Gweagal peoples of the Tharawal nation. 

%%%%%%%%%%%%%%%%%%%%%%%%%%%%%%%%%%%%%%%%%%%%%%%%%%



% \bibliographystyle{mnras}
% \bibliography{references} % if your bibtex file is called example.bib

%%%%%%%%%%%%%%%%%%%%%%%%%%%%%%%%%%%%%%%%%%%%%%%%%%

%%%%%%%%%%%%%%%%% APPENDICES %%%%%%%%%%%%%%%%%%%%%

%\pagebreak

\appendix

\section{Frequency analysis}\label{apx:freqs}

The frequencies present in the time series of each star were determined by a standard prewhitening procedure as described by \citep{papics17} down to a signal-to-noise ratio (SNR) of four. The SNR was calculated in a 3-d$^{-1}$ wide window during prewhitening. The extracted frequencies, amplitudes and phases are given in Tables~\ref{tab:Alcyone}~to~\ref{tab:Pleione}. While these frequencies have not been corrected for stellar line-of-sight Doppler velocity shifts as advocated by \citet{davies14}, the radial velocities of these stars are only 5--8\,km\,s$^{-1}$, so the differences are negligible. The zero-point of the time-scale for the phases is BJD 2454833.

The frequency resolution of each time series (the Rayleigh limit) is 1/$T$ = 0.014d$^{-1}$, where $T$ is the length of the time series. Frequencies that were found within the Rayleigh limit of a higher SNR peak have been excluded. We further note that several of the remaining frequencies are within the \citet{loumos78} criterion of 1.5$\times$ the Rayleigh limit, which means their Fourier parameters will have been affected by the prewhitening procedure. Low-amplitude, low-frequency peaks that may be an artifact of the long-term trend removal have also been excluded. Possible combination frequencies have been indicated, although it is not possible to be sure of the correct identification given the closely-spaced frequencies and limited resolution, and the base frequencies may not necessarily have the highest amplitudes \citep[see][]{kurtz15}. 

\newpage

\begin{table}
	\centering
	\caption{Fourier parameters of the Alcyone time series}
	\label{tab:Alcyone}
	\begin{tabular}{ccccr} % four columns, alignment for each
		\hline
		ID & Frequency & Amplitude & Phase & SNR\\
		   & (d$^{-1}$) & (ppm) & (rad) & \\
		\hline
	    $\nu_{1}\sim\nu_{4}-\nu_{3}$  &  0.1482$\pm$0.0009 &  63$\pm$7 & \;\;2.2$\pm$0.7 &  7.6 \\
		$\nu_{2}$  &  0.4215$\pm$0.0009 &  49$\pm$6 & \;\;0.9$\pm$0.7 &  7.1 \\
		$\nu_{3}$  &  0.4360$\pm$0.0004 & 151$\pm$8 & $-$2.9$\pm$0.3 & 17.5 \\
		$\nu_{4}$  &  0.5798$\pm$0.0014 &  18$\pm$3 & $-$2.4$\pm$1.1 &  4.0 \\
		$\nu_{5}$  &  0.7452$\pm$0.0013 &  25$\pm$4 & $-$0.5$\pm$1.0 &  5.0 \\
		$\nu_{6}$  &  0.7922$\pm$0.0012 &  36$\pm$5 & $-$0.5$\pm$0.9 &  6.6 \\
		$\nu_{7}$  &  0.8084$\pm$0.0012 &  28$\pm$4 & $-$1.0$\pm$1.0 &  5.2 \\
		$\nu_{8}$  &  0.8266$\pm$0.0012 &  24$\pm$7 & $-$0.2$\pm$1.0 &  4.5 \\
		$\nu_{9}\sim2\nu_{3}$  &  0.8729$\pm$0.0011 &  35$\pm$5 & \;\;0.8$\pm$0.9 &  5.8 \\
		$\nu_{10}\sim3\nu_{3}$  &  1.3051$\pm$0.0017 &  14$\pm$3 & $-$2.5$\pm$1.4 &  4.2 \\
		$\nu_{11}\sim2\nu_{8}$ &  1.6480$\pm$0.0019 &  18$\pm$4 & $-$1.8$\pm$1.5 &  4.9 \\
		$\nu_{12}\sim\nu_{6}+\nu_{9}$ &  1.6648$\pm$0.0011 &  34$\pm$5 & $-$1.8$\pm$0.9 &  5.5 \\
		$\nu_{13}\sim4\nu_{2}$ &  1.6871$\pm$0.0013 &  38$\pm$6 & \;\;0.7$\pm$1.0 &  6.9 \\
		$\nu_{14}\sim4\nu_{3}$ &  1.7483$\pm$0.0019 &  13$\pm$3 & $-$0.2$\pm$1.6 &  4.1 \\
		$\nu_{15}$ &   3.453$\pm$0.003  &   7$\pm$3 & $-$3.0$\pm$2.6 &  4.2 \\
		$\nu_{16}$ &  23.731$\pm$0.005  &   5$\pm$4 & \;\;1.3$\pm$4.3 &  4.4 \\
		\hline
	\end{tabular}
\end{table}

\begin{table}
	\centering
	\caption{Fourier parameters of the Atlas time series}
	\label{tab:Atlas}
	\begin{tabular}{ccccr} % four columns, alignment for each
		\hline
		ID & Frequency & Amplitude & Phase & SNR \\
		   & (d$^{-1}$) & (ppm) & (rad) & \\
		\hline
        $\nu_{1}$             & 0.4119$\pm$0.0005 & 2110$\pm$130 & $-$1.0$\pm$0.4 & 23.2 \\
        $\nu_{2}=\nu_{6} - 2 \nu_{1}$ & 0.5092$\pm$0.0008 &   68$\pm$7   & $-$2.5$\pm$0.7 &  6.4 \\
        $\nu_{3}$             & 0.7627$\pm$0.0008 &  143$\pm$16  & $-$2.4$\pm$0.7 &  9.4 \\
        $\nu_{4}=2\nu_{1}$            & 0.8237$\pm$0.0001 & 1128$\pm$18  & \;\;0.6$\pm$0.1 & 30.3 \\
        $\nu_{5}=3\nu_{1}$            & 1.2365$\pm$0.0009 &   80$\pm$9   & \;\;2.2$\pm$0.7 &  6.8 \\
        $\nu_{6}$             & 1.3406$\pm$0.0005 & 1500$\pm$100 & $-$1.5$\pm$0.4 & 25.5 \\
        $\nu_{7}=2\nu_{3}$            & 1.5252$\pm$0.0013 &   43$\pm$7   & \;\;0.6$\pm$1.0 &  4.2 \\
        $\nu_{8}=4\nu_{1}$            & 1.6504$\pm$0.0015 &   36$\pm$7   & $-$0.5$\pm$1.2 &  4.3 \\
        $\nu_{9}=2\nu_{6}$            & 2.6827$\pm$0.0016 &   62$\pm$13  & \;\;0.5$\pm$1.3 &  8.0 \\
        
		\hline
	\end{tabular}
\end{table}

\begin{table}
	\centering
	\caption{Fourier parameters of the Electra time series}
	\label{tab:Electra}
	\begin{tabular}{ccccr} % four columns, alignment for each
		\hline
		ID & Frequency & Amplitude & Phase & SNR \\
		   & (d$^{-1}$) & (ppm) & (rad) & \\
		\hline
        $\nu_{1}$ & 0.8584$\pm$0.0009 & 38$\pm$4 & \;\;2.0$\pm$0.7 &  6.6 \\
        $\nu_{2}$ & 0.9031$\pm$0.0006 & 58$\pm$4 & \;\;0.4$\pm$0.5 & 10.0 \\
        $\nu_{3}$ & 0.9632$\pm$0.0013 & 17$\pm$3 & \;\;0.2$\pm$1.0 &  3.9 \\
        $\nu_{4}\sim2\nu_{2}$ & 1.7986$\pm$0.0011 & 24$\pm$4 &  $-$1.5$\pm$0.9 &  5.2 \\
        $\nu_{5}\sim2\nu_{3}$ & 1.9183$\pm$0.0013 & 18$\pm$3 &  $-$2.2$\pm$1.0 &  4.2 \\
        $\nu_{6}\sim2\nu_{3}$ & 1.9377$\pm$0.0010 & 32$\pm$4 &  $-$1.9$\pm$0.8 &  5.6 \\
        $\nu_{7}\sim\nu_{2}+2\nu_{3}-\nu_{1}$ & 1.9842$\pm$0.0013 & 19$\pm$3 &  $-$2.0$\pm$1.0 &  4.5 \\
		\hline
	\end{tabular}
\end{table}

\begin{table}
	\centering
	\caption{Fourier parameters of the Maia time series}
	\label{tab:Maia}
	\begin{tabular}{ccccr} % four columns, alignment for each
		\hline
		ID & Frequency & Amplitude & Phase & SNR \\
		   & (d$^{-1}$) & (ppm) & (rad) & \\
		\hline
		$\nu_{1}$  & 0.0972$\pm$0.0002 & 1300$\pm$40 &  $-$0.79$\pm$0.17 & 34.0 \\
		\hline
	\end{tabular}
\end{table}

\begin{table}
	\centering
	\caption{Fourier parameters of the Merope time series}
	\label{tab:Merope}
	\begin{tabular}{ccccr} % four columns, alignment for each
		\hline
		ID & Frequency & Amplitude & Phase & SNR \\
		   & (d$^{-1}$) & (ppm) & (rad) & \\
		\hline
		$\nu_{1}$  & 0.9082$\pm$0.0012 &   76$\pm$11  & $-$0.7$\pm$0.9 &  5.7 \\
		$\nu_{2}$  & 2.0132$\pm$0.0007 &  166$\pm$14  & $-$3.1$\pm$0.5 &  8.5 \\
		$\nu_{3}$  & 2.0263$\pm$0.0004 &  323$\pm$18  & \;\;2.0$\pm$0.3 & 11.9 \\
		$\nu_{4}$  & 2.0487$\pm$0.0004 & 2030$\pm$100 & \;\;2.0$\pm$0.3 & 23.7 \\
		$\nu_{5}$  & 2.0721$\pm$0.0006 &  204$\pm$16  & \;\;0.2$\pm$0.5 & 10.2 \\
		$\nu_{6}$  & 2.0855$\pm$0.0010 &   92$\pm$12  & \;\;0.5$\pm$0.8 &  5.9 \\
		$\nu_{7}\sim2\nu_{4}$ & 4.0834$\pm$0.0007 &  264$\pm$23  & $-$2.4$\pm$0.5 & 15.2 \\
		$\nu_{8}\sim3\nu_{4}$ & 6.1300$\pm$0.0015 &   76$\pm$14  & \;\;0.7$\pm$1.2 &  9.2 \\
		\hline
	\end{tabular}
\end{table}

\begin{table}
	\centering
	\caption{Fourier parameters of the Taygeta time series}
	\label{tab:Taygeta}
	\begin{tabular}{ccccr} % four columns, alignment for each
		\hline
		ID & Frequency & Amplitude & Phase & SNR \\
		   & (d$^{-1}$) & (ppm) & (rad) & \\
		\hline
        $\nu_{1}$  &  0.4038$\pm$0.0013 &  35$\pm$6  & $-$1.6$\pm$1.1 &  4.2  \\
        $\nu_{2}$  &  0.4184$\pm$0.0009 &  91$\pm$10 & $-$1.4$\pm$0.7 &  7.8  \\
        $\nu_{3}$  &  0.4449$\pm$0.0008 & 115$\pm$12 & \;\;1.9$\pm$0.6 &  9.1  \\
        $\nu_{4}$  &  0.4795$\pm$0.0012 &  40$\pm$6  & $-$1.1$\pm$1.0 &  4.3  \\
        $\nu_{5}$  &  0.5577$\pm$0.0011 &  61$\pm$8  & $-$2.9$\pm$0.9 &  5.6  \\
        $\nu_{6}$  &  0.6384$\pm$0.0007 & 147$\pm$14 & \;\;2.9$\pm$0.6 & 10.3  \\
        $\nu_{7}$  &  0.7618$\pm$0.0014 &  35$\pm$6  & $-$2.7$\pm$1.1 &  4.2  \\
        $\nu_{8}\sim\nu_{5}+\nu_{6}$  &  1.1952$\pm$0.0014 &  52$\pm$9  & \;\;1.0$\pm$1.1 &  5.5  \\
        $\nu_{9}\sim3\nu_{6}-\nu_{5}$  &  1.3557$\pm$0.0010 &  90$\pm$11 & \;\;0.5$\pm$0.8 &  7.4  \\
        $\nu_{10}$ & 19.8724$\pm$0.0042 &  12$\pm$7  & $-$1.5$\pm$3.4 &  4.4  \\
		\hline
	\end{tabular}
\end{table}

\begin{table}
	\centering
	\caption{Fourier parameters of the Pleione time series}
	\label{tab:Pleione}
	\begin{tabular}{ccccr} % four columns, alignment for each
		\hline
		ID & Frequency & Amplitude & Phase & SNR \\
		   & (d$^{-1}$) & (ppm) & (rad) & \\
		\hline
		$\nu_{1}$ & 0.3200$\pm$0.0014 &  190$\pm$30 & \;\;1.8$\pm$ 1.1 &  4.6 \\
		$\nu_{2}$ & 0.7352$\pm$0.0014 &  170$\pm$30 &  $-$0.8$\pm$ 1.2 &  4.1 \\
		$\nu_{3}$ & 1.7101$\pm$0.0012 &  250$\pm$40 &  $-$1.8$\pm$ 1.0 &  5.0 \\
		$\nu_{4}$ & 1.7324$\pm$0.0008 &  520$\pm$50 &  $-$0.3$\pm$ 0.6 & 10.9 \\
		$\nu_{5}$ & 1.7605$\pm$0.0004 & 1160$\pm$60 & \;\;2.3$\pm$ 0.3 & 18.4 \\
		$\nu_{6}$ & 7.4599$\pm$0.0053 &   50$\pm$40 & \;\;0.0$\pm$ 4.2 &  4.5 \\		
		\hline
	\end{tabular}
\end{table}

%%%%%%%%%%%%%%%%%%%%%%%%%%%%%%%%%%%%%%%%%%%%%%%%%%

% Don't change these lines
\bsp	% typesetting comment
\label{lastpage}
\end{document}